\documentclass[12pt,letter]{article}

\usepackage{graphicx,  color}
\def\eslt{E_T^{\rm miss}}
\def\to{\rightarrow}

\def\bi{\begin{itemize}}
\def\ei{\end{itemize}}
\def\te{\tilde e}
\def\ta{\tilde a}

\def\tu{\tilde u}

\def\tb{\tilde b}

\def\tst{\tilde t}

\def\ttau{\tilde \tau}
\def\tmu{\tilde \mu}
\def\tg{\tilde g}
\def\tnu{\tilde\nu}
\def\tell{\tilde\ell}
\def\tq{\tilde q}
\def\tw{\widetilde W}
\def\tz{\widetilde Z}
\def\alt{\stackrel{<}{\sim}}
\def\agt{\stackrel{>}{\sim}}
\def\be{\begin{equation}}  
\def\ee{\end{equation}}  
\def\bea{\begin{eqnarray}}  
\def\eea{\end{eqnarray}}  
\def\beas{\begin{eqnarray*}}  
\def\eeas{\end{eqnarray*}}  
\newcommand\prd[3]{{\it Phys.\ Rev.\ }{\bf D #1} (#2) #3}
\newcommand\prep[3]{{\it Phys.\ Rept.\ }{\bf #1} (#2) #3}
\newcommand\prl[3]{{\it Phys.\ Rev.\ Lett.\ }{\bf #1} (#2) #3}
\newcommand\plb[3]{{\it Phys.\ Lett.\ }{\bf B #1} (#2) #3}
\newcommand\jhep[3]{{\it J. High Energy Phys.\ }{\bf #1} (#2) #3}

\newcommand\npb[3]{{\it Nucl.\ Phys.\ }{\bf B #1} (#2) #3}
\newcommand\epjc[3]{{\it Eur.\ Phys.\ J. }{\bf C #1} (#2) #3}

\newcommand\zpc[3]{{\it Z.\ Phys.\ {\bf C}}{\bf  #1} (#2) #3}

\newcommand{\hepph}[1]{hep-ph/#1}

\newcommand{\hepex}[1]{hep-ex/#1}

\begin{document}
\begin{titlepage}

\vspace{0.5cm}
\begin{center}
{\Large \bf 
Gaugino Anomaly Mediated SUSY Breaking:\\ 
phenomenology and prospects for the LHC
}\\ 
\vspace{1.2cm} \renewcommand{\thefootnote}{\fnsymbol{footnote}}
{\large Howard Baer$^{1}$\footnote[1]{Email: baer@nhn.ou.edu },
Senarath de Alwis$^2$\footnote[2]{Email: dealwiss@colorado.edu}, 
Kevin Givens$^2$\footnote[3]{Email: kevin.givens@colorado.edu},\\
Shibi Rajagopalan$^1$\footnote[3]{Email: shibi@nhn.ou.edu},
Heaya Summy$^1$\footnote[4]{Email: heaya@nhn.ou.edu}} \\
\vspace{1.2cm} \renewcommand{\thefootnote}{\arabic{footnote}}
{\it 
1. Dept. of Physics and Astronomy,
University of Oklahoma, Norman, OK 73019, USA \\
2. Dept. of Physics,
University of Colorado, Boulder CO 80309, USA \\
}

\end{center}

\vspace{0.5cm}
\begin{abstract}
\noindent 
We examine the supersymmetry phenomenology of a novel scenario of supersymmetry (SUSY) 
breaking which we call Gaugino Anomaly Mediation, or inoAMSB. 
This is suggested by recent work on the phenomenology of flux compactified type IIB string theory. 
The essential features of this scenario are that the  gaugino masses are of the 
anomaly-mediated SUSY breaking (AMSB) form, while
scalar and trilinear soft SUSY breaking terms are highly suppressed. 
Renormalization group effects yield an allowable sparticle mass spectrum, 
while at the same time avoiding charged LSPs; the latter are common in models with 
negligible soft scalar masses, such as no-scale or gaugino mediation models. 
Since scalar and trilinear soft terms are highly suppressed, the 
SUSY induced flavor and $CP$-violating processes are also suppressed.
The lightest SUSY particle is the neutral wino, while the heaviest
is the gluino. In this model, there should be a strong multi-jet $+\eslt$ signal 
from squark pair production at the LHC.
We find a 100 fb$^{-1}$ reach of LHC out to $m_{3/2}\sim 118$ TeV, corresponding to
a gluino mass of $\sim 2.6$ TeV.
A double mass edge from the opposite-sign/same flavor dilepton invariant mass distribution 
should be visible at LHC; this, along with the presence of short-- but visible-- highly ionizing
tracks from quasi-stable charginos, should provide a smoking gun signature for inoAMSB.
\vspace*{0.8cm}


\end{abstract}


\end{titlepage}

\section{Introduction}
\label{sec:intro}

String theory is very attractive in that it allows for a consistent quantum mechanical
treatment of gravitation, while at the same time including all the necessary
ingredients for containing the well-known gauge theories which comprise the
Standard Model of particle physics. 
Phenomenologically viable versions of string theory require the stabilization
of all moduli fields as well as weak to intermediate scale supersymmetry
breaking. Models satisfying these criteria were first developed in
the context of type IIB string theory using flux compactifications
and non-perturbative effects 
on Calabi Yau orientifolds (CYO's) (for reviews see \cite{Grana:2005jc} \cite{Douglas:2006es}). 
The low energy limit of type-IIB string theory after compactification on a CYO
is expected to be $N=1$ supergravity (SUGRA).

Two classes of the above models which yield an interesting supersymmetry breaking scenario have been studied:
\begin{itemize}
\item [{a)}] Those with only a single K\"ahler modulus (SKM models). These are  essentially
of the KKLT type \cite{Kachru:2003aw} but with uplift coming from one-loop quantum effects. 
\item [{b)}] Large Volume Scenario (LVS)\cite{Balasubramanian:2005zx}
models which require at least two moduli. 
\end{itemize}
In both of these types of models, the moduli fields are stabilized
using a combination of fluxes and non-perturbative effects. Additionally,
supersymmetry is broken by the moduli fields acquiring non-zero F-terms
and interacting gravitationally with the MSSM. For both models, the
gauginos acquire mass predominately through a Weyl anomaly effect
while the classical contribution to the scalar masses and trilinear
coupling constants are naturally suppressed. 

Generically in a string model there are three types of contributions to the soft SUSY breaking terms: 
\begin{enumerate}
\item Terms generated by classical string theory effects. 
\item Terms generated by  quantum effects (effectively string loop
corrections).
\item Weyl anomaly (AMSB) contributions\cite{amsb} to the gaugino masses.
\end{enumerate}
In this class of models, the MSSM may be  located on  D3 branes at a singularity 
or on D7 branes wrapping a collapsing four cycle in the CYO.  
The string theory calculations are expected to give boundary conditions at (or near) the string scale
 $M_{string}$, which
may range (almost) up to the 
GUT scale  or as low as some
intermediate scale $\ll M_{GUT}$. In both cases, the classical string theory as
well as 1-loop quantum contributions to the soft SUSY breaking terms are suppressed
relative to the weak scale. For gaugino masses, it has been shown that these will 
gain contributions from the Weyl anomaly, and therefore assume the usual form
as expected in models with anomaly-mediated SUSY breaking:
\be
M_i= \frac{b_i g_i^2}{16\pi^2}m_{3/2}.
\ee
Here $i$ labels the gauge group, $g_i$ is the associated gauge coupling, $m_{3/2}$ is
the gravitino mass, and $b_i$ is the co-efficient of the gauge group beta function:
$b_i=(33/5,1,-3)$.
Meanwhile, soft SUSY breaking scalar masses will have generically suppressed 
classical string masses and 1-loop contributions, and  receive 
{\it no contribution} from Weyl anomalies\cite{deAlwis:2008aq}. 

To good approximation, we can set in this class of models,
\be
m_0=A_0=0, 
\ee
where $m_0$ is the common soft SUSY breaking scalar mass at $M_{string}$ and 
$A_0$ is the trilinear soft SUSY breaking (SSB) term. 

In addition,  the bilinear SSB mass $B$ and the superpotential
$\mu$ term would be zero at the classical (AdS) minimum. 
However these can acquire non-zero values once uplift terms that correct the value of the cosmological constant are turned on.
Here, we will feign ignorance as to the origin of these terms, and instead
adopt a phenomenological approach which determines their values by 
finding an appropriate minimum of the electroweak scalar potential. The minimization 
procedure allows one to trade the $B$ parameter for the ratio of Higgs field vevs $\tan\beta$, and 
to require the value of $|\mu |$ which is needed in order to specify the correct mass 
of the $Z$ boson\cite{wss}.
In this case, the SKM and LVS models would both be well-described by the following parameter space
\be
m_{3/2},\ \tan\beta ,\ sign(\mu ) ,
\label{eq:pspace}
\ee
where in addition we take $m_0=A_0=0$. 
While the SSB scalar and trilinear terms are small at $M_{string}$, 
they can become large at $Q=M_{weak}$ due to renormalization group (RG) running.

In fact, these sort of RG boundary conditions are similar to those of
no-scale supergravity models (NS)\cite{noscale}, 
and also gaugino-mediated SUSY breaking (inoMSB) models\cite{inomsb}.
However, in both NS and inoMSB models, it is expected that the gaugino masses unify
to a common gaugino mass $m_{1/2}$ at the string scale. 
The fact that both scalar and trilinear soft SUSY breaking terms have only small
contributions at the high scale (compactification scale $M_c$, $M_{P}$ or $M_{GUT}$) 
is highly desirable for solving the
SUSY flavor and $CP$ problems. In the general MSSM, unconstrained off-diagonal
terms in the scalar and trilinear sector lead to large contributions to
flavor changing and $CP$ violating processes, for which there are tight limits\cite{masiero}.
Under renormalization group evolution, the off-diagonal terms remain small, 
while diagonal terms receive significant contributions due to gauge interactions and the large
gaugino masses.

Using the NS or inoMSB boundary conditions, it is well known that one gains a
sparticle mass spectrum with $\tau$ sleptons as the lightest SUSY particle (LSP)\footnote{
One way out is to hypothesize the gravitino as LSP. In our case, we will find that the
gravitino mass is always in the multi-TeV range.}.
In models with $R$-parity conservation, the $\ttau_1$ would be absolutely stable, 
thus violating constraints coming from search experiments for long lived, stable charged 
relics from the Big Bang. 
One way around this dilemma has been suggested by Schmaltz and Skiba\cite{ss}:
adopting $M_{string}> M_{GUT}$, so that above-the-GUT-scale running lifts the value
of $m_0$ above zero at the GUT scale.
Another possibility is to allow for unconstrained, or a less-constrained, form of
{\it non-universal} gaugino masses\cite{derm_mafi}.

We find here that the $m_0\sim A_0\sim 0$ boundary conditions-- along with the 
AMSB form  for gaugino masses-- in fact leads to viable sparticle mass spectra across most of parameter
space, without the need for above-the-GUT-scale running, or a less-constrained form for
gaugino masses, or an artifically light gravitino mass.
While these boundary conditions seem to emerge naturally in type IIB string models 
with flux compactifications, we may also consider such boundary conditions by themselves
as being perhaps more general, and well-motivated by their desirable low energy features.
For this reason, we will hereafter refer to the class of models leading to the above
boundary conditions 
as gaugino anomaly mediated SUSY breaking, or {\it inoAMSB} models, 
since only the gaugino masses receive 
contributions of the AMSB form, and the other soft parameters are similar to those as 
generated in gaugino mediation.

The remainder of this paper is organized as follows. In Sec. \ref{sec:model}, 
we review some of the string theoretic supergravity model details that motivate us to
consider the inoAMSB form of boundary conditions. 
In Sec. \ref{sec:pspace}, we plot out the spectra of
superpartners that is expected in the inoAMSB model. While some features are
similar to what is known as ``minimal'' AMSB (mAMSB), some crucial differences exist that may allow one
to distinguish inoAMSB from mAMSB and also ``hypercharged'' anomaly-mediation (HCAMSB)\cite{hcamsb}. 
We also examine what happens if the string scale is taken to be some intermediate value, 
or if some small universal scalar mass is adopted.
We also plot out low energy constraints from $BF(b\to s\gamma )$ and $(g-2)_\mu$. 
In Sec. \ref{sec:lhc}, we examine the sort of signatures expected from the
inoAMSB model at the CERN LHC. 
Since squark masses are always lighter than gluinos, we expect a large rate for
$\tq\tq$ and $\tq\tg$ production, leading to a large rate for multi-jet$+\eslt$ events.
Since the lightest SUSY particle is a neutral wino, as in most AMSB-type models, we
expect a nearly mass degenerate, quasi-stable chargino, which can lead to short but observable
highly ionizing tracks in a collider detector.
In addition, squarks cascade decay to neutralinos, followed by neutralino decay to
lepton plus either left- or right-slepton states. The unique cascade decay pattern leads to a 
distinct double mass edge in the same-flavor/opposite-sign dilepton invariant mass 
distribution, which distinguishes inoAMSB from mAMSB or HCAMSB. 
In Sec. \ref{sec:conclude}, we present our conclusions.

\section{Effective supergravity from IIB strings: Overview of Models}
\label{sec:model}

\subsection{Effective Supergravity Theory}

The low energy limit of IIB string theory, after compactification on
a Calabi-Yao orientifold, yields ${\cal N}=1$ supergravity.
The (superspace) action then has the generic form (see for example
\cite{Wess:1992cp,Gates:1983nr}) 
\begin{eqnarray}
S & = & -3\int d^{8}z{\bf E}\exp[-\frac{1}{3}K(\Phi,\bar{\Phi};C,\bar{C}e^{2V})]+\nonumber \\
 &  & \left(\int d^{6}z2{\cal E}[W(\Phi,C)+\frac{1}{4}f_{a}(\Phi){\cal W}^{a}{\cal W}^{a}]
+h.c.\right) ,\label{eq:sugraaction}
\end{eqnarray}
where we have set $M_{P}=(8\pi G_{N})^{-1/2}=2.4\times10^{18}GeV=1$.
Here $K$-- the K\"ahler potential-- is a real superfield as is the gauge
pre-potential $V$.  $W$-- the superpotential-- is a holomorphic field,
as is the gauge coupling function $f_a$ and the (fermionic) gauge (super)
field strength ${\cal W}(V)$. ${\bf E}d^{8}z$ is the full superspace
measure and ${\cal E}d^{6}z$ is the chiral superspace measure. Ignoring
the $D$-terms, which are zero at the minimum of the potential in
the class of models considered here, the SUGRA potential takes the
standard form (after going to the Einstein frame)\textcolor{black}
{\begin{equation}
{\color{blue}{\color{black}V(\Phi)=F^{A}F^{\bar{B}}K_{A\bar{B}}-3|m_{3/2}(\Phi)|^{2}}} .\label{eq:potential}
\end{equation}
Here $F^{A}=e^{K/2}K^{A\bar{B}}D_{\bar{B}}W,\, D_{A}W\equiv\partial_{A}W+K_{A}W$
where $K_{A}=\partial_{A}K,\, K_{A\bar{B}}=\partial_{A}\partial_{\bar{B}}K$
and ${\color{blue}|{\color{black}m_{3/2}|^{2}\equiv e^{K}|W|^{2}}}$
becomes the squared gravitino mass when evaluated at the minimum of
the potential.}

We separate the chiral superfields of the theory into moduli fields
(which come from string theory and describe the internal geometry
of the CY manifold ) and the dilaton (which are collectively called
$\Phi$) and the MSSM fields (which we have called $C$). Expanding
$K$ and $W$ in powers of the MSSM fields we have:

\begin{eqnarray}
W & = & \hat{W}(\Phi)+\mu(\Phi)H_{d}H_{u}+\frac{1}{6}Y_{\alpha\beta\gamma}(\Phi)C^{\alpha}C^{\beta}C^{\gamma}+\ldots,\label{eq:W}\\
K & = & \hat{K}(\Phi,\bar{\Phi})+\tilde{K}_{\alpha\bar{\beta}}(\Phi,\bar{\Phi})C^{\alpha}C^{\bar{\beta}}+[Z(\Phi,\bar{\Phi})H_{d}H_{u}+h.c.]+\ldots\label{eq:K}\\
f_{a} & = & f_{a}(\Phi).\label{eq:f_a}\end{eqnarray}
Here we have separated the two Higgs superfield multiplets ($H_{d,u})$.
The moduli fields essentially play the role of spurion fields that
break supersymmetry, once they are stabilized and acquire a definite
vacuum expectation value such that one or more of them also has a
non-zero F-term.

Also
\begin{eqnarray}
\hat{K} & = & -2\ln\left({\cal V}+\frac{\hat{\xi}}{2}\right)-\ln\left(i\int\Omega\wedge\bar{\Omega}(U,\bar{U})\right)-\ln(S+\bar{S}),\ \ {\rm and}  \label{eq:hatK}\\
\hat{W} & = & \int G_{3}\wedge\Omega+\sum_{i}A_{i}e^{-a_{i}T^{i}}.\label{eq:hatW}
\end{eqnarray}
Here ${\cal V}$ is the volume of the CYO and is a function of the
K\"ahler moduli superfields $T^{i}$ (with $i=1,\cdots , h_{11}$), and $\hat{\xi}$
is a stringy ($\alpha'$) correction that is an $O(1)$ number depending
on the Euler character of the CYO and the real part of the dilaton
superfield $S$. $\Omega$ is the holomorphic three form on the CYO
and is a function of the complex structure moduli superfields $U_{r}$ (with $r=1,\cdots,h_{21}$).

\subsection{Single K\"ahler modulus scenario}

In this construction, type IIB string theory is compactified on a
CYO and the MSSM lives on a stack of D3 branes.
We take a CYO with just one K\"ahler modulus, $T$, (i.e. $h_{11}=1$)
but with a number $\sim 10^{2}$ of complex moduli, $U_{r}$. These moduli,
along with the axio-dilaton, $S$, are then stabilized using internal
fluxes and non-perturbative effects. Classically, we can find a minimum
of this potential with the $F$-term of $T$ being $\sim m_{3/2}$
with the other moduli $F$-terms being suppressed. The cosmological constant
would be negative but suppressed. The soft terms are also highly suppressed.
This solution of course receives quantum mechanical corrections starting at 1-loop.
In terms of an effective field theory description, they would depend
on a string scale cutoff $\Lambda$. These can serve to uplift the
cosmological constant and to generate the soft SUSY breaking masses, proportional
to $\frac{\Lambda}{4\pi}m_{3/2}$. The cutoff $\Lambda$ is essentially
the string scale and in this class of models may be taken as large
as $10^{-2}M_{P}$ so that a GUT scenario could be accommodated. This class of models 
is discussed in \cite{deAlwis:2008kt}.

\subsection{Large Volume Scenario (LVS)}

In this class of models\cite{Balasubramanian:2005zx}, we again
compactify IIB string theory on a CYO. However,
we now consider more than one K\"ahler modulus, $T^{i}(i=1,\cdots , h_{11})$.
In particular-- in the simplest such situation-- we have a large modulus,
$\tau_{b}$, and small moduli, ($\tau_{s}$, $\tau_{a}$), controlling
the overall size of the CYO and the volume of two small  4-cycles respectively.
The total volume is then given by 
\begin{equation}
\mathcal{V}=\tau_{b}^{3/2}-\tau_{s}^{3/2}-\tau_{a}^{3/2} .
\end{equation}
This is referred to as a {}``Swiss Cheese'' model. Again the MSSM
may be located on D3 branes at a singularity. Alternatively, we could
have it on a stack of D7 branes wrapping a four cycle (taken to be
the one labelled by the index $a$). In this case, it has been argued
\cite{Blumenhagen:2007sm,Blumenhagen:2009gk} that the necessity
of having chiral fermions on this brane prevents this cycle from being
stabilized by non-perturbative effects and it shrinks below the string scale. 
Effectively, this means that the physics is the same as in the D3 brane case.

 Extremizing the potential leads to an exponentially large volume\cite{Balasubramanian:2005zx} 
${\cal V}\sim e^{a\tau_{s}},\,\tau_{s}\sim\hat{\xi}$.
It turns out that the suppression of FCNC effects lead to 
$\mathcal{V}\agt 10^{5}\mathit{l}_{P}^{6}$\cite{deAlwis:2009tp} (where $l_P$ is the Planck length), 
so the string scale is $M_{string}\alt M_{P}/\sqrt{{\cal V}}\sim10^{15.5}{\rm GeV}$.
The minimum of the potential (CC) is given by 
$V_{0}\sim-\frac{m_{3/2}^{2}M_P^2}{\ln m_{3/2}\mathcal{V}}$.
This minimum can be uplifted to zero when $S$ and $U_r$ acquire (squared)
$F$-terms of the order $\frac{m_{3/2}^{2}M_P^2}{\ln m_{3/2}\mathcal{V}}$.
Classical contributions to the scalar and slepton masses are also
of this same order. With the above lower bound on the volume, this
means that even for $m_{3/2}\sim 100$ TeV,  the classical soft terms
are $\alt 100$ GeV. Of course if one wants to avoid fine-tuning
of the flux superpotential, we would need to take even larger values
of ${\cal V}$ corresponding to a string scale of $10^{12}$ GeV. In
this case the classical soft terms are completely negligible (for
$m_{3/2}\sim100$ TeV) but  the (classical) $\mu$-term is also strongly suppressed. 

In the rest of this section  we will call the holomorphic variable associated with the large modulus $\tau_b$, $T$.

\subsection{Gaugino Masses - Weyl Anomaly Effects}

For a generic version of supergravity, the gaugino masses satisfy
the following relation at the classical level: 
\begin{equation}
\frac{M_{a}}{g_{a}^{2}}=\frac{1}{2}F^{A}\partial_{A}f_{a}(\Phi) .
\end{equation}
For both the single K\"ahler modulus model and LVS cases, the leading
contribution to the gauge coupling function $f_{a}(\Phi)$ comes from
the axio-dilaton $S$, so at a classical minimum where the SUSY breaking
is expected to be in the $T$ modulus direction, 
the string theoretic contribution to the  gaugino mass is highly suppressed.

However, there is an additional contribution to the gaugino mass due
to the (super) Weyl anomaly. This comes from the expression for the
effective gauge coupling superfield that has been derived by Kaplunovsky
and Louis \cite{Kaplunovsky:1994fg} (KL)%
\footnote{As explained in \cite{deAlwis:2008aq}, the usual formulae for AMSB
\cite{amsb}, \cite{Bagger:1999rd} need modification
in the light of \cite{Kaplunovsky:1994fg}.%
}. 
For the gaugino masses, the relevant contribution comes from taking
the $F$-term of 
\begin{equation}
H_{a}(\Phi,\tau,\tau_{Z})=f_{a}(\Phi)-\frac{3c_{a}}{8\pi^{2}}\ln\phi-\frac{T_{a}(r)}{4\pi^{2}}\phi_{Z}.\label{eq:H}
\end{equation}
Here, the first term on the RHS is the classical term; the second comes
from the anomaly associated with rotating to the Einstein-K\"ahler
frame.  $c_{a}=T(G_{a})-\sum_rT_{a}(r)$ is the anomaly coefficient and the last term comes from
the anomaly associated with the transformation to canonical kinetic
terms for the MSSM fields. Also note that we have ignored the gauge
kinetic term normalization anomaly \cite{ArkaniHamed:1997mj,deAlwis:2008aq}
which is a higher order effect. The chiral superfields $\phi, \phi_r$ that generate
these transformations are given by, 
\begin{eqnarray}
\ln\phi+\ln\bar{\phi} & = & \frac{1}{3}K|_{\rm Harm} ,\\
\phi_{r}+\bar{\phi}_{r} & = & \ln\det\tilde{K}_{\alpha\bar{\beta}}^{(r)} .
\end{eqnarray}
The instruction on the RHS of the first equation is to take the sum of the chiral and anti-chiral (i.e. harmonic) part of the expression.
 After projecting the appropriate $F$ terms we arrive at the following
expression: 
\begin{equation}
\frac{2M_{a}}{g_{a}^{2}}=F^{A}\partial_{A}f_{a}-\frac{c_{i}}{8\pi^{2}}F^{A}K_{A}-\sum_{r}\frac{T_{i}(r)}{4\pi^{2}}F^{A}\partial_{A}\ln\det\tilde{K}_{\alpha\bar{\beta}}^{(r)} .
\end{equation}
 As pointed out earlier, the first (classical) term is greatly suppressed
relative to $m_{3/2}$. The dominant contribution therefore comes
from the last two (Weyl anomaly) contributions. It turns out that
(after using the formulae $F^{T}=-(T+\bar{T})m_{3/2}$, $K_{T}=-3/(T+\bar{T})$
and $\tilde{K}_{\alpha\bar{\beta}}=k_{\alpha\beta}/(T+\bar{T})$ which
are valid up to volume suppressed corrections), this yields%
\footnote{Note that we expect the Weyl anomaly expressions for the gaugino masses
given below to be valid only because of the particular (extended no-scale)
features of this class of string theory models. It so happens that
these are exactly the same as the expressions given in what is usually
called AMSB: but that is an accident due entirely to the fact that
in these extended no-scale models the relationship $F^{A}K_{A}\simeq3m_{3/2}$
is true.%
}, \begin{equation}
M_{a}=\frac{b_{a}g_{a}^{2}}{16\pi^{2}}m_{3/2},\label{eq:weylgaugino}\end{equation} where $b_a=-3T(G_{a})+\sum_rT_{a}(r)$ is the beta function coefficient.

\subsection{Scalar Masses, Trilinear Couplings, $\mu$ and $B\mu$ terms}

Here we summarize the results from this class of string theory models for
the values of the  soft parameters at the UV scale, 
{\it i.e.} $\Lambda\sim M_{string}\sim M_P/\sqrt{{\cal V}}$. 
These values should be the initial conditions for the RG evolution of these parameters. 
In the LVS case, it was estimated\cite{deAlwis:2009tp} that the lower bound on the CYO volume was 
${\cal V}>10^5$. Also, we will choose typical values  $h_{21}\sim O(10^2)$ for the number of complex structure moduli. 
We will also take the gravitino mass $m_{3/2}\sim |W|M_P/{\cal V}\sim 50\ {\rm TeV}$.
Such a large value of $m_{3/2}$ allows us to avoid the SUGRA gravitino problem, which leads
to a disruption of Big Bang nucleosynthesis if $m_{3/2}\alt 5$ TeV and $T_R\agt 10^5$ GeV\cite{moroi}.

Unlike the gaugino masses, scalar masses and trilinear soft terms
do not acquire corrections from the Weyl anomaly. They are essentially
given at the UV scale by their classical string theory value plus one loop string/effective field
theory corrections. In the $h_{11}=1$ case, 
the classical soft terms are essentially zero while in the LVS case 
\begin{equation}
m_{0}\sim O\left(\frac{m_{3/2}}{\sqrt{\ln m_{3/2}{\cal V}}}\right),\,\mu\sim\frac{B\mu}{\mu}\alt \sqrt{h_{21}}m_{0},\, A_0\ll m_{0}.\label{eq:lvsclassical} 
\end{equation}

After adding quantum corrections at the UV scale both cases give similar values for the soft terms.  
As an example, we illustrate for two values for the CYO volume:   
\begin{itemize}
\item 
${\cal V}\sim 10^5, M_{string}\sim\Lambda\sim10^{-2.5}M_P\sim 10^{15.5}\ {\rm GeV}$.  
Then,  
\begin{equation}\mu \sim\frac{B\mu}{\mu}\alt 250\ {\rm GeV},\, m_{0}\sim 25\ {\rm GeV},
\,A_0\ll m_0 .\label{GUT}\end{equation}

\item  ${\cal V}\sim 10^{12},\ M_{string}\sim\Lambda\sim10^{-6}M_P\sim 10^{12}\ {\rm GeV}$. 
Then, 
\begin{equation}\mu \sim\frac{B\mu}{\mu}\alt 10^{-1}\ {\rm GeV},
\, m_{0}\sim 10^{-2}\ {\rm GeV},\, A_0\ll m_0 .\label{INT}
\end{equation}
The second very large volume case can be accessed only in the LVS model.

\end{itemize}
The first case is at the lower bound for the volume. 
This gives the largest allowable string scale. 
This is still somewhat below the apparent unification scale,  but it is close enough that 
(allowing for undetermined $O(1)$ factors) we may use the GUT scale as the point at which to 
impose the boundary conditions. This is useful for the purpose of comparing with other models 
of SUSY mediation where it is conventional to use the GUT scale.
    
The second case above corresponds to choosing generic values of the flux superpotential, 
while the first needs a fine tuned set of fluxes to get $|W|\sim 10^{-8}$, 
in order to have a gravitino mass of  $\sim 10^2$ TeV, though in type IIB string theory general arguments show that there exist a large number of solutions which allow this. 
The most significant problem with the second case 
(apart from the fact that there is no hope of getting a GUT scenario) is 
 the extremely  low upper bound on the $\mu$ term. 
In other words, there is a serious $\mu$- problem. 
The first case also may have a $\mu$ term problem, but again since these estimates are accurate 
only to $O(1)$ numbers, 
it is possible to envisage that the problem can be resolved within the context of this model.

In any case, as we discussed in the introduction, 
we are going to take an approach where the string theory input is used to suggest a 
class of phenomenological models. Given that in both the GUT scale model and the 
intermediate scale model, the soft scalar mass and $A$ term are suppressed well below the weak scale, 
we will input the value zero for these at the UV scale, 
while the gaugino masses at this scale are given by (\ref{eq:weylgaugino}).

We also discuss the case when the input scalar mass $m_0$ is non-negligible. 
This would be the case for instance in the SKM model with smaller volumes and/or larger values of $h_{21}$,  and also in the case of LVS with the volume at the 
lower bound but with larger values of $h_{21}$. 

\section{Mass spectra, parameter space and constraints for the inoAMSB model}
\label{sec:pspace}

\subsection{Sparticle mass spectra and parameter space}

We begin our discussion by examining the sort of sparticle mass spectra that is
expected from  the inoAMSB boundary conditions: $m_0=A_0=0$ but 
with $M_i=\frac{b_i g_i^2}{16\pi^2}m_{3/2}$. 
We compute the sparticle mass spectra using the Isasugra subprogram of the
event generator Isajet\cite{isajet}, along with the option of
non-universal gaugino masses. 
The parameter space is that of Eq. \ref{eq:pspace}.

After input of the above parameter set, Isasugra
implements an iterative procedure of solving the MSSM RGEs for the
26 coupled renormalization group equations, taking the weak scale 
measured gauge couplings and third generation Yukawa couplings as inputs, as well
as the above-listed GUT scale SSB terms. Isasugra implements full 2-loop RG running
in the $\overline{DR}$ scheme, and minimizes the RG-improved 1-loop effective
potential at an optimized scale choice $Q=\sqrt{m_{\tst_L}m_{\tst_R}}$
(which accounts for leading two-loop terms)\cite{hh} 
to determine the magnitude of $\mu$ and the value of $m_A$. All physical sparticle masses
are computed with complete 1-loop corrections, and 1-loop weak scale threshold corrections
are implemented for the $t$, $b$ and $\tau$ Yukawa couplings\cite{pbmz}. The off-set of the 
weak scale boundary conditions due to threshold corrections (which depend on the entire
superparticle mass spectrum), necessitates an iterative up-down RG running solution.
The resulting superparticle mass spectrum is typically in close accord with other
sparticle spectrum generators\cite{kraml}.

\begin{figure}[htbp]
\begin{center}
\includegraphics[width=0.5\textwidth]{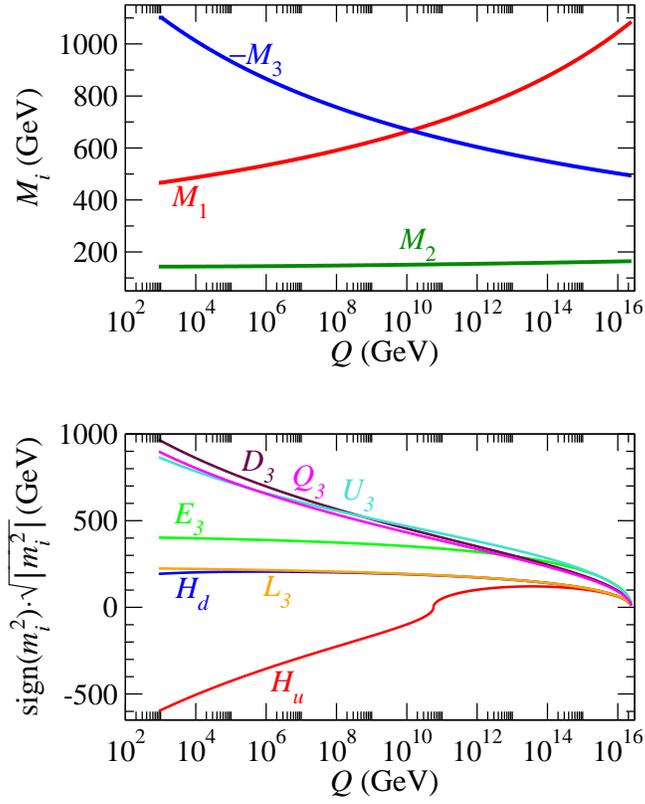}
\caption{Running of soft SUSY breaking parameters as a function of
energy scale $Q$ for $m_{3/2}=50$ TeV, $\tan\beta =10$ and $\mu >0$ 
in the inoAMSB model, with $M_{string}=M_{GUT}$.
}
\label{fig:run}
\end{center}
\end{figure}

We begin by examining a single point in inoAMSB parameter space, 
where $m_{3/2}=50$ TeV, and $\tan\beta =10$, with $\mu >0$ as suggested by
the $(g-2)_\mu$ anomaly\cite{brown,isagm2}. 
In Fig. \ref{fig:run}, we plot in frame {\it a}). the running gaugino masses, and
in frame {\it b}). the running third generation and Higgs soft SUSY breaking scalar 
masses. We actually plot here $sign(m_i^2)\times \sqrt{|m_i^2|}$, in order to show
the possible running to negative squared masses, while at the same time showing the
true scale of the soft terms in GeV units. Frame {\it a}). is as expected in most AMSB
masses {\it i.e.} where $M_1\gg |M_3|\gg M_2$ at $Q=M_{string}$, where here we take 
$M_{string}= M_{GUT}$. The RG running of the gaugino masses leads to $-M_3\gg M_1$
at $Q=M_{weak}$, while $M_2$ remains the lightest of gaugino masses at the weak
scale, leading to a wino-like lightest neutralino $\tz_1$, which might also be the
lightest SUSY particle (LSP). In frame {\it b})., we see that the SSB scalar masses,
beginning with negligible GUT scale values, are initially pulled up to positive values,
mainly by the influence of the large value of $M_1$ at the GUT scale. In fact, the
right-slepton mass $m_{E_3}^2$ initially evolves to the highest values, since it has the largest
hypercharge quantum number $Y=2$. The disparate $Y$ values between $E_3$ and the doublet $L_3$
lead ultimately to a large splitting between left- and right- slepton SSB masses in the
inoAMSB case, while these masses tend to be quite degenerate in mAMSB\cite{amsb_lhc}. As the scale
$Q$ moves to values $\ll M_{GUT}$, QCD effects pull the squarks to much higher masses:
in this case around the TeV scale, while sleptons, which receive no QCD contribution, remain in
the 200-400 GeV range. The value of $m_{H_u}^2$ is driven as usual to negative squared values, 
resulting in a radiative breakdown of electroweak symmetry (REWSB). Since $M_2<\sqrt{m_{L_3}^2}$, we
generically find a wino-like neutralino as the LSP, and there is no problem with a charged
LSP (as in NS/inoMSB models) or tachyonic sleptons (as in AMSB).

\begin{figure}[htbp]
\begin{center}
\includegraphics[angle=-90,width=0.7\textwidth]{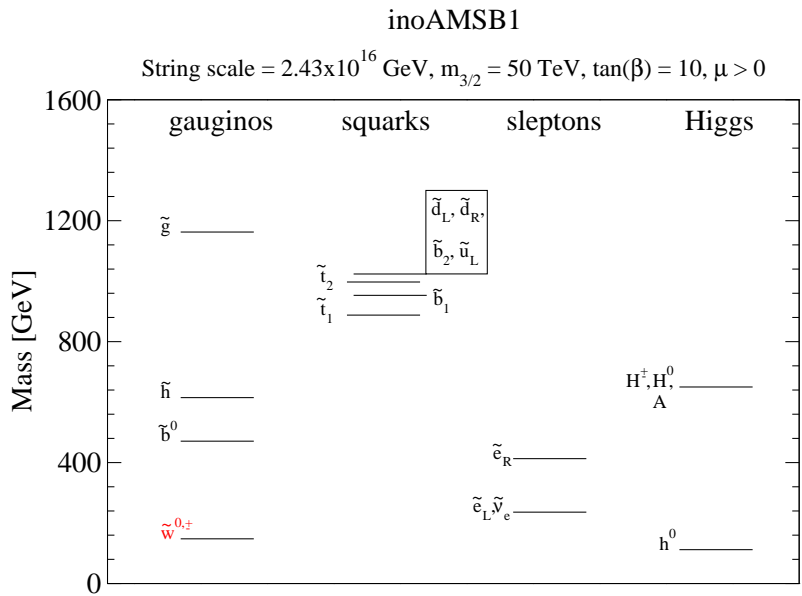}
\caption{Plot of sparticle masses for the inoAMSB1 case study
with $m_{3/2}=50$ TeV, $\tan\beta =10$ and $\mu >0$.
}
\label{fig:BMs}
\end{center}
\end{figure}

Once the weak scale SSB terms are computed, then the physical mass eigenstates
and mixings may be computed, and one-loop mass corrections added. The resulting physical mass spectrum
is listed schematically in Fig. \ref{fig:BMs}{\it a}). and in Table \ref{tab:BMs}, column 3.
We adopt this inoAMSB model as a benchmark case, labeled inoAMSB1.
In Table \ref{tab:BMs}, we also show for comparison two related cases with 
$m_{3/2}=50$ TeV and $\tan\beta =10$: for mAMSB supersymmetry in
column 1, with $m_0=300$ GeV, and in HCAMSB\cite{hcamsb}, column 2, with
mixing parameter $\alpha =0.025$.\footnote{In the HCAMSB model, while most of the MSSM resides 
on a visible brane, $U(1)$ gauginos propagate in the bulk. Thus, the SSB 
boundary conditions, taken at the GUT scale, are those of AMSB, but with an additional
contribution to the hypercharge gaugino mass, proportional to the mixing parameter $\alpha$.}
While the first three cases listed in Table \ref{tab:BMs} 
have similar
values of $m_{\tg}$ and $m_{\tw_1,\tz_1}$ (due to the same input value of $m_{3/2}$), 
we see that inoAMSB1 has the previously noted large $\te_L$-$\te_R$ splitting, with 
$m_{\te_L}<m_{\te_R}$, while
mAMSB has nearly degenerate $\te_L$ and $\te_R$, with $m_{\te_R}<m_{\te_L}$. However, the left-right 
slepton splitting in inoAMSB1 is not as severe as that shown in HCAMSB1, from Ref. \cite{hcamsb},
where an even larger value of $M_1$ at $M_{GUT}$ is expected. In the HCAMSB1 case, the
$\tz_4$ state tends to be nearly pure bino-like, whereas in inoAMSB1, it is instead higgsino-like.
%

%
\begin{table}
\begin{center}
\begin{tabular}{lcccc}
\hline
parameter & mAMSB & HCAMSB1 & inoAMSB1 & inoAMSB2 \\
\hline
$\alpha$    & --- & 0.025 & --- & ---  \\
$m_0$       & 300 & --- & --- & --- \\
$m_{3/2}$   & $50\ {\rm TeV}$ & $50\ {\rm TeV}$ & $50\ {\rm TeV}$ & $100\ {\rm TeV}$ \\
$\tan\beta$ & 10 & 10 & 10 & 10\\
$M_1$       & 460.3   & 997.7 & 465.5 & 956.1 \\
$M_2$       & 140.0   & 139.5 & 143.8 & 287.9 \\
$\mu$       & 872.8 & 841.8 & 607.8 & 1127.5 \\
$m_{\tg}$   & 1109.2 & 1107.6 & 1151.0 & 2186.1 \\
$m_{\tu_L}$ & 1078.2 & 1041.3 & 1011.7 & 1908.7 \\
$m_{\tu_R}$ & 1086.2   & 1160.3 & 1045.1 & 1975.7 \\
$m_{\tst_1}$& 774.9 & 840.9 & 878.8 & 1691.8 \\
$m_{\tst_2}$& 985.3 & 983.3 & 988.4 & 1814.8 \\
$m_{\tb_1}$ & 944.4 & 902.6 & 943.9 & 1779.5  \\
$m_{\tb_2}$ & 1076.7 & 1065.7 & 1013.7 & 1908.3 \\
$m_{\te_L}$ & 226.9 & 326.3  & 233.7 & 457.8 \\
$m_{\te_R}$ & 204.6 & 732.3  & 408.6 & 809.5 \\
$m_{\tw_2}$ & 879.2 & 849.4 & 621.2 & 1129.8 \\
$m_{\tw_1}$ & 143.9 & 143.5 & 145.4 & 299.7 \\
$m_{\tz_4}$ & 878.7 & 993.7 & 624.7 & 1143.2 \\ 
$m_{\tz_3}$ & 875.3 & 845.5 & 614.4 & 1135.8 \\ 
$m_{\tz_2}$ & 451.1 & 839.2 & 452.6 & 936.8 \\ 
$m_{\tz_1}$ & 143.7 & 143.3 & 145.1 & 299.4 \\ 
$m_A$       & 878.1 & 879.6 & 642.9 & 1208.9 \\
$m_h$       & 113.8 & 113.4 & 112.0 & 116.0 \\ 
$\Omega_{\tz_1}h^2$ & 0.0016 & 0.0015 & 0.0016 & 0.007 \\
\hline
$\sigma\ [{\rm fb}]$ & $7.7\times 10^3$ & $7.4\times 10^3$ & $7.5\times 10^3$ & $439$ \\
$\tg ,\tq\ {\rm pairs}$ & 15.0\% & 15.5\% & 19.1\% & 3\% \\
${\rm EW-ino\  pairs}$ & 79.7\% & 81.9\% & 75.6\% & 93\% \\
${\rm slep.}\ {\rm pairs}$ & 3.7\% & 0.8\% & 3.1\% & 3\% \\
$\tst_1\bar{\tst}_1$ & 0.4\% & 0.2\% & 0.1\% & 0\% \\
\hline
\end{tabular}
\caption{Masses and parameters in~GeV units
for four case study points mAMSB1, HCAMSB1, inoAMSB1 and inoAMSB2
using Isajet 7.80 with $m_t=172.6$ GeV and $\mu >0$. 
We also list the 
total tree level sparticle production cross section 
in fb at the LHC.
}
\label{tab:BMs}
\end{center}
\end{table}

\begin{figure}[htbp]
\begin{center}
\includegraphics[angle=-90,width=0.75\textwidth]{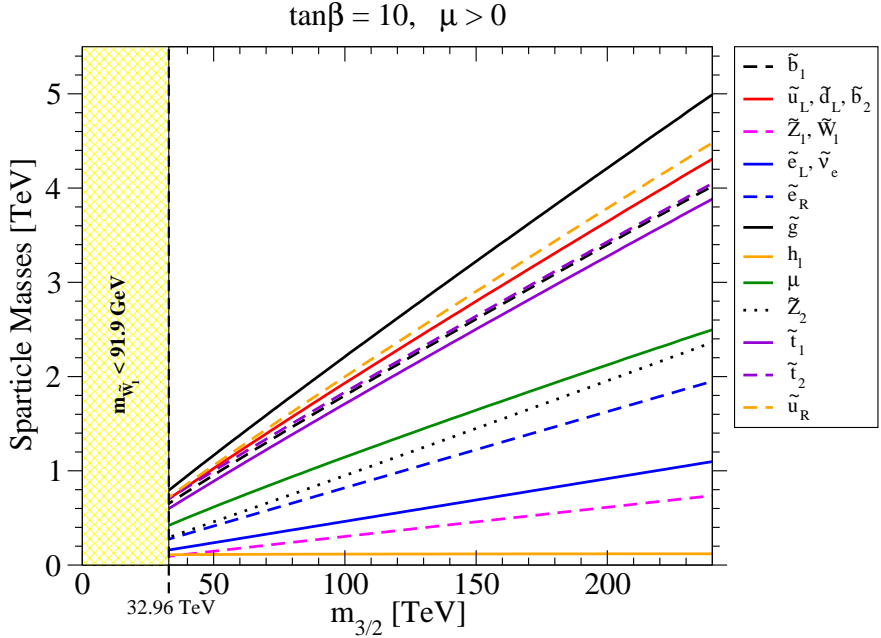}
\vspace{-.5cm}
\caption{Sparticle mass spectrum versus $m_{3/2}$ in 
the inoAMSB with $M_{string}=M_{GUT}$, $\tan\beta =10$, with $\mu >0$ and
$m_t=172.6$ GeV. 
}
\label{fig:m10}
\end{center}
\end{figure}

Next, we investigate the effect of varying $m_{3/2}$ on the sparticle
mass spectrum.
We plot in Fig. \ref{fig:m10} the mass spectra of
various sparticles versus $m_{3/2}$ in the inoAMSB model
while taking $\tan\beta =10$, $\mu >0$ and $m_t=172.6$ GeV.
The lowest value of $m_{3/2}$ which is allowed is $m_{3/2}=32.96$ TeV.
Below this value, $m_{\tw_1}<91.9$ GeV, which is excluded by LEP2 searches
for charginos from AMSB models\cite{lepw1lim}. We see from Fig. \ref{fig:m10}
that there is a characteristic mass hierarchy in the inoAMSB model, 
where $m_{\tz_1,\tw_1}<m_{\te_L, \tnu_{eL}}<m_{\te_R}<|\mu |<m_{\tq}<m_{\tg}$.
As $m_{3/2}$ increases, all these masses grow, but the relative hierarchy is maintained.
For such a spectrum with $m_{\tq}<m_{\tg}$ and with relatively light sleptons, 
we would thus expect LHC collider events which are dominated by squark pair production, followed by 
squark cascade decays $\tq\to q\tz_i\to q \tell^\pm\ell^\mp$, which would lead to events with 
two hard jets (plus additional softer jets) and rich in isolated leptons coming from 
cascade decay produced sleptons.

\begin{figure}[htbp]
\begin{center}
\includegraphics[angle=-90,width=0.75\textwidth]{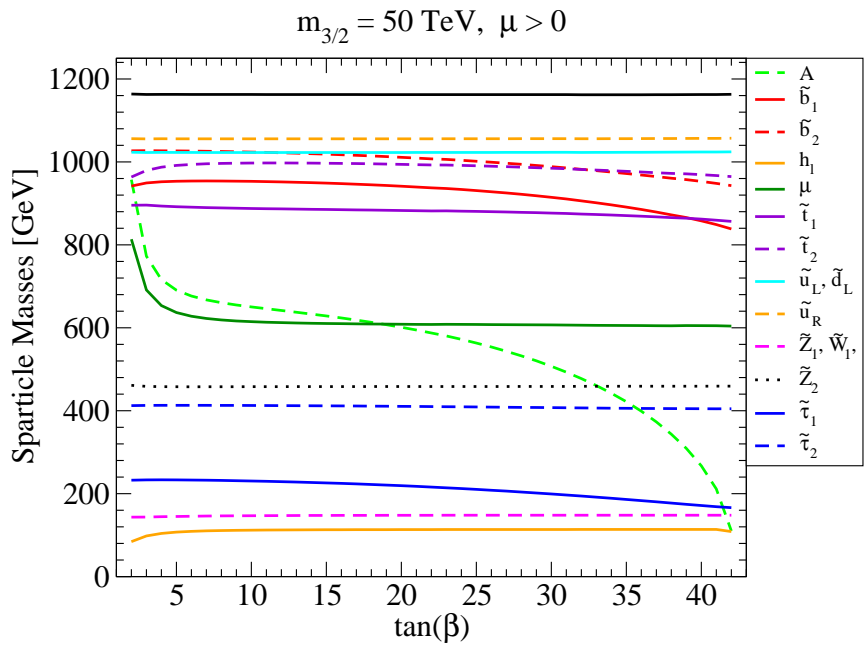}
\vspace{-.5cm}
\caption{Sparticle mass spectrum versus $\tan\beta$ for $m_{3/2}=50$ TeV in 
the inoAMSB with $M_{string}=M_{GUT}$ and with $\mu >0$.
}
\label{fig:mtanb}
\end{center}
\end{figure}

In Fig. \ref{fig:mtanb}, we show the variation in sparticle masses against
$\tan\beta$ with $m_{3/2}$ fixed at 50 TeV. As $\tan\beta$ increases, the
$b$ and $\tau$ Yukawa couplings both increase. These act to suppress the sbottom and stau
SSB mass terms, and also give larger left-right mixing to the mass eigenstates.
In addition, the value of $m_{H_d}^2$ is pushed to negative values by the large $b$ and 
$\tau$ Yukawa couplings. The value of $m_A$ is given approximately from the 
EWSB minimization conditions as $m_A^2\sim m_{H_d}^2-m_{H_u}^2$.
Since the mass gap between $m_{H_u}^2$ and $m_{H_d}^2$ drops as $\tan\beta$ increases, 
the value of $m_A$ also drops sharply with increasing $\tan\beta$. 
The point at which $m_A$ drops below limits from LEP2 searches 
(and shortly thereafter REWSB no longer occurs
in  a valid fashion) provides the high $\tan\beta$ boundary to the parameter space.

\begin{figure}[htbp]
\begin{center}
\includegraphics[angle=-90,width=0.7\textwidth]{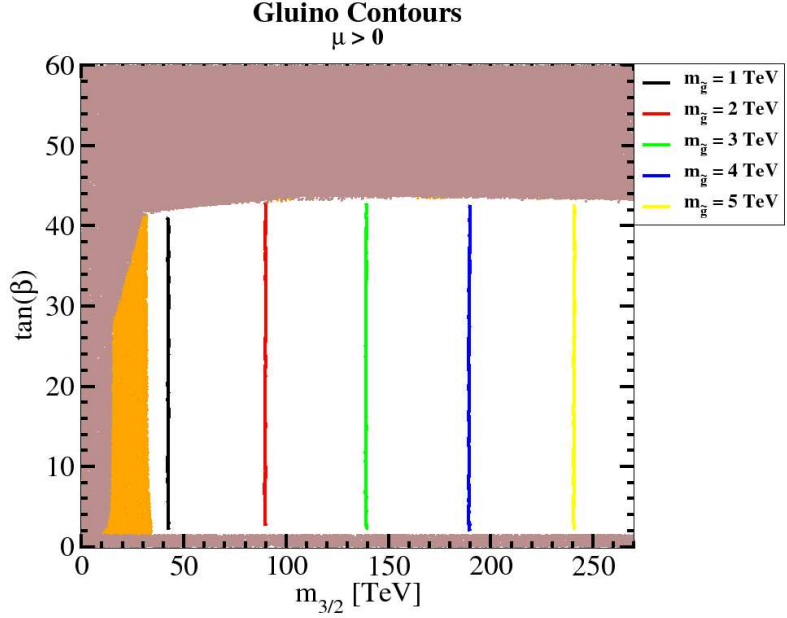}
\caption{Allowed parameter space of the inoAMSB models in the
$m_{3/2}$ vs. $\tan\beta$ plane with  $\mu >0$.
We plot also contours of $m_{\tg}$. 
}
\label{fig:pspace}
\end{center}
\end{figure}

In Fig. \ref{fig:pspace}, we show the entire parameter space for
the inoAMSB model in the $m_{3/2}$ vs. $\tan\beta$ plane for $\mu >0$
with $m_t=172.6$ GeV. The gray shaded region gives allowable sparticle mass spectra. 
The orange region gives chargino masses below the
LEP2 limit, and so is experimentally excluded. 
The brown shaded region for $\tan\beta \agt 42$ is excluded because REWSB no longer
occurs in an appropriate fashion.
The brown shaded region at very low $\tan\beta$
gives too light a value of $m_h$: here we require $m_h>111$ GeV (even though LEP2
requires $m_h>114.4$ GeV), due to a projected $\sim \pm 3$ GeV theory error on our
lightest Higgs mass calculation. 
We also show contours of $m_{\tg}$ ranging from 1-5 TeV.  The $m_{\tg}\sim 5$ TeV range 
will surely be beyond the reach of LHC.

\begin{figure}[htbp]
\begin{center}
\includegraphics[angle=-90,width=0.7\textwidth]{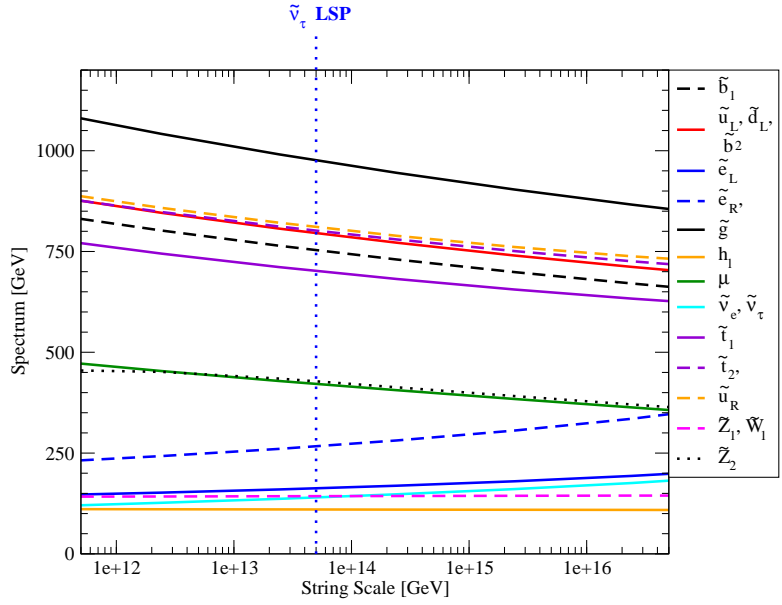}
\caption{Plot of sparticle masses for the inoAMSB1 case study
with $m_{3/2}=50$ TeV, $\tan\beta =10$ and $\mu >0$,
but with variable value of $M_{string}$.
}
\label{fig:Mstring}
\end{center}
\end{figure}

As noted in Sec. \ref{sec:model}, in the inoAMSB model the string scale
$M_{string}$ need not be equal to $M_{GUT}$. If it is not, then
it can have significant effects on the sparticle mass spectrum.
The sparticle mass spectrum versus variable $M_{string}$ is shown in Fig. \ref{fig:Mstring}
for the case where $m_{3/2}=50$ TeV, $\tan\beta =10$ and $\mu >0$. Here, we see 
that the sparticle mass spectrum spreads out as $M_{string}$ varies from
$M_{GUT}$ down to $10^{11}$ GeV. In addition, some important level crossings occur.
Most important of these is that for $M_s\alt 5\times 10^{13}$ GeV, the $\tnu_\tau$ state
becomes the lightest MSSM particle, and for even lower $M_s$ values,
$m_{\tnu_e}$ and $m_{\tnu_\mu}$ drop below $m_{\tz_1}$.
There already exist severe limits on stable sneutrino dark matter\cite{sneu_dm},
which discourage this type of scenario. If we insist upon a neutralino as LSP, 
then we must take not too low a value of $M_s$.

\begin{figure}[htbp]
\begin{center}
\includegraphics[angle=-90,width=0.7\textwidth]{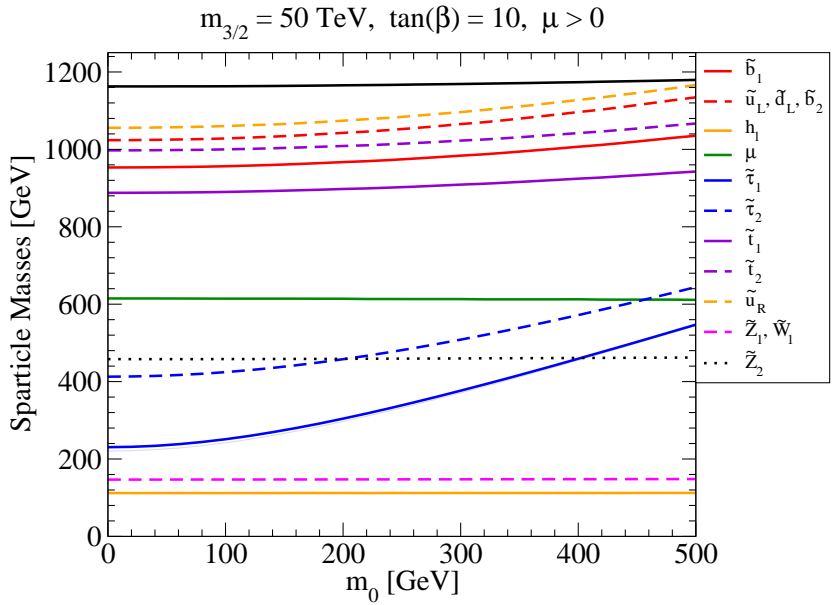}
\caption{Plot of sparticle masses for the inoAMSB 
with $m_{3/2}=50$ TeV, $\tan\beta =10$ and $\mu >0$,
but with an additional universal contribution $m_0$ added to all scalar masses.
}
\label{fig:m0}
\end{center}
\end{figure}

Finally, we note that in the inoAMSB model, scalar masses and $A$-parameters
are expected to be suppressed, but they are not expected to be exactly zero.
In Fig. \ref{fig:m0}, we show the mass spectra from inoAMSB models where we
add an additional universal mass contribution $m_0$ to all scalars.
We adopt values $m_{3/2}=50$ TeV and $\tan\beta =10$ for this plot. 
As $m_0$ increases beyond zero, it is seen that the spectra change little so long as
$m_0\alt 100$ GeV, and also the mass orderings remain intact. For larger values of 
$m_0$, the left- and right- slepton masses begin to increase, with first
$m_{\te_R}$ surpassing $m_{\tz_2}$, and later even $m_{\te_L}$ surpasses 
$m_{\tz_2}$. At these high values of $m_0$, decay modes such as $\tz_2\to\ell^\pm\tell^\mp$
would become kinematically closed, thus greatly altering the collider signatures.
However, generically in this class of models, we would not expect such large
additional contributions to scalar masses. 

\subsection{$BF(b\to s\gamma )$ and $(g-2)_\mu$ in inoAMSB}

Along with experimental constraints on the inoAMSB models from LEP2 limits on $m_h$ and $m_{\tw_1}$, 
there also exist indirect limits on model parameter space from
comparing measured values of $BF(b\to s\gamma )$ and 
$\Delta a_\mu\equiv(g-2)_\mu /2$ against SUSY model predictions.

\subsubsection{$BF(b\to s\gamma )$}

\begin{figure}[htbp]
\begin{center}
\includegraphics[angle=270,width=0.8\textwidth]{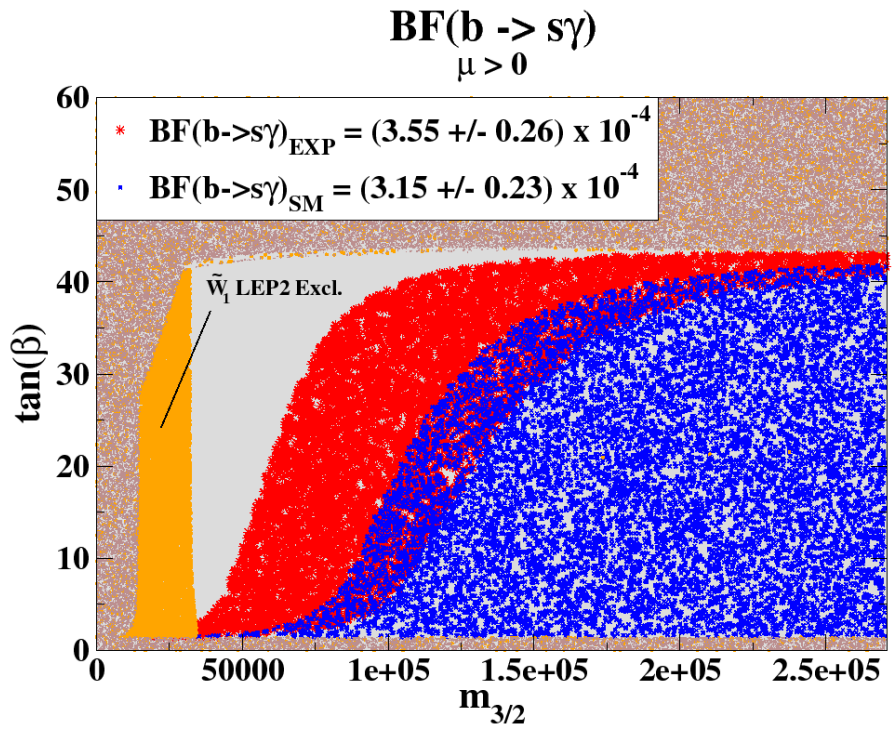}
\caption{Branching fraction for $b\to s\gamma$ versus $m_{3/2}$ 
and $\tan\beta$ variation in the inoAMSB model with $M_{string}=M_{GUT}$.
}
\label{fig:bsg}
\end{center}
\end{figure}

As an example, we show in Fig. \ref{fig:bsg} regions of the branching fraction
for $BF(b\to s\gamma )$ in the inoAMSB model versus $m_{3/2}$ and $\tan\beta$ 
variation, calculated using the Isatools subroutine 
ISABSG\cite{isabsg}. The red-shaded region corresponds to branching fraction values
within the SM theoretically predicted region $BF(b\to s\gamma )_{SM}=(3.15\pm 0.23)\times 10^{-4}$, 
by a recent evaluation by Misiak\cite{misiak}). 
The blue-shaded region corresponds to branching fraction values within the
experimentally allowed region\cite{bsg_ex}: here, 
the branching fraction $BF(b\to s\gamma )$ has been measured by the
CLEO, Belle and BABAR collaborations; a combined analysis\cite{bsg_ex}
finds the branching fraction to be $BF(b\to s\gamma )=(3.55\pm                  
0.26)\times 10^{-4}$.
The gray shaded region gives too large a value of $BF(b\to s\gamma )$. This region occurs
for low $m_{3/2}$, where rather light $\tst_1$ and $\tw_1$ lead to large branching fractions, or
large $\tan\beta$, where also the SUSY loop contributions are enhanced\cite{bbct}.

\subsubsection{$(g-2)_\mu /2$}

\begin{figure}[htbp]
\begin{center}
\includegraphics[angle=-90,width=0.8\textwidth]{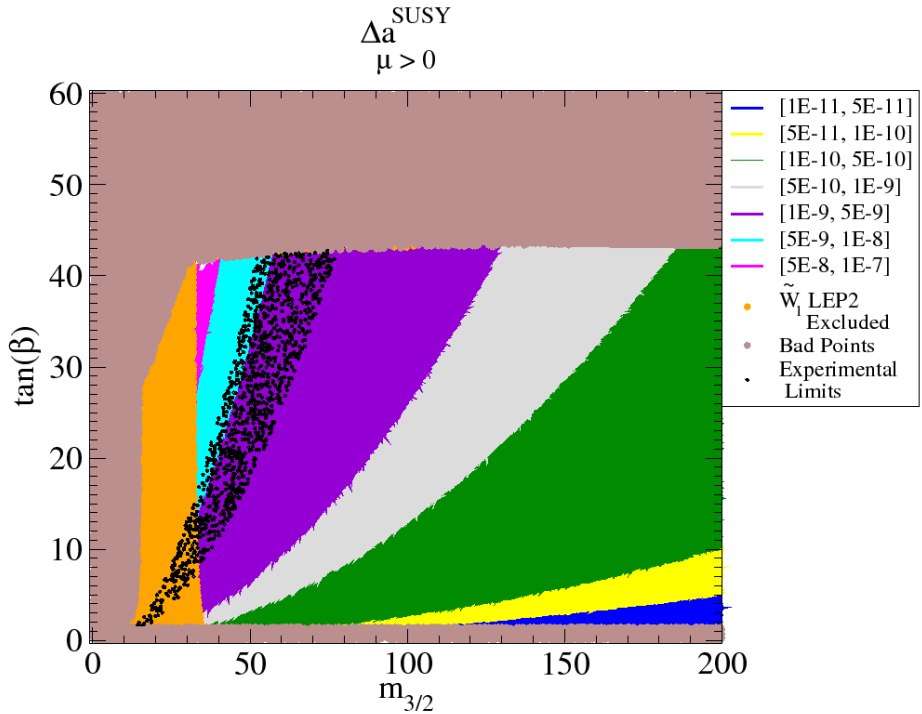}
\caption{SUSY contribution to $\Delta a_\mu$  versus $m_{3/2}$
and $\tan\beta$ variation in the inoAMSB model with $M_{string}=M_{GUT}$.
We also take $\mu >0$ and $m_t=172.6$ GeV.
}
\label{fig:gm2}
\end{center}
\end{figure}

In Fig. \ref{fig:gm2}, we plot the SUSY contribution to $\Delta a_\mu$:
$\Delta a_\mu^{SUSY}$ (using ISAAMU from Isatools\cite{isagm2}). 
The contribution is large when $m_{3/2}$ is small;
in this case, rather light $\tmu_L$ and $\tnu_{\mu L}$ masses lead to
large deviations from the SM prediction. 
The SUSY contributions to $\Delta a_\mu^{SUSY}$ also increase with $\tan\beta$. 
It is well-known that there
is a discrepancy between the SM predictions for $\Delta a_\mu$, where
$\tau$ decay data, used to estimate the hadronic vacuum polarization 
contribution to $\Delta a_\mu$, gives rough accord with the SM, while
use of $e^+e^-\to hadrons$ data at very low energy leads to a 
roughly $3\sigma$ discrepancy.
The measured $\Delta a_\mu$ anomaly, given as $(4.3\pm 1.6)\times 10^{-9}$ by the 
Muon $g-2$ Collaboration\cite{brown}, is shown by the black dotted region.

\subsubsection{Dark matter in inoAMSB}

Finally, we remark upon the relic density of dark matter in the inoAMSB model.
If thermal production of the lightest neutralino is assumed to give the 
dominant DM in the universe, then all over parameter space, the predicted
neutralino abundance $\Omega_{\tz_1}h^2$ is far below the WMAP 
measured value of $\Omega_{CDM}h^2\sim 0.11$. 
Some sample calculated values are listed in Table \ref{tab:BMs}.
It has been suggested in Ref. \cite{randallmoroi} that
production and decay of moduli fields or other processes can also contribute
to the DM abundance. 
Decay of moduli fields in the early universe could then account for the
discrepancy between the measured DM abundance and the predicted thermal
abundance in inoAMSB models.

As an alternative, if the strong $CP$ problem is solved via the 
Peccei-Quinn mechanism, then a superfield containing the axion/axino multiplet
should occur. In this case, a mixture of axions\cite{axion} and axinos\cite{axino}, 
rather than wino-like neutralinos, could constitute the DM abundance\cite{mix}. 
The exact abundance will depend on the axino mass $m_{\ta}$, the Peccei-Quinn
breaking scale $f_a$, and the re-heat temperature $T_R$ after inflation.

In light of these two alternative DM mechanisms, we regard the 
inoAMSB parameter space as essentially unconstrained by the measured
abundance of DM in the universe.

\section{The inoAMSB model and the LHC}
\label{sec:lhc}

\subsection{Sparticle production at LHC}

In the inoAMSB model, for benchmark point inoAMSB1, we list several
sparticle production cross sections in Table \ref{tab:BMs}.
We see that for this case, the dominant sparticle production consists
of electroweak-ino pair production: mainly 
$pp\to \tw_1^+\tw_1^-\ {\rm and}\ \tw_1^\pm\tz_1$ reactions.
Since $\tz_1$ is stable (or quasi-stable in the event of light axino dark matter), 
and mainly $\tw_1^\pm\to\pi^\pm\tz_1$
(where the $\pi^\pm$ is very soft), these reactions do not provide enough 
visible energy to meet detector trigger requirements (unless there is 
substantial initial state radiation).
 
\begin{figure}[htbp]
\begin{center}
\includegraphics[angle=-90,width=0.8\textwidth]{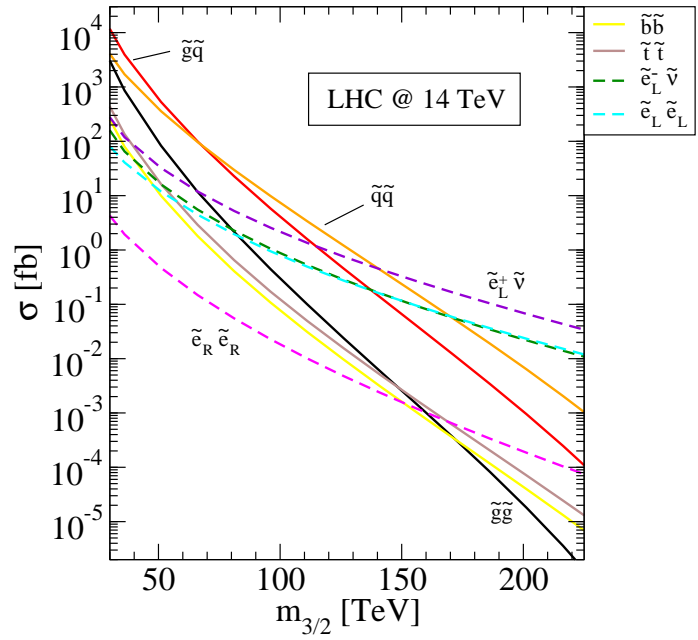}
\caption{Sparticle pair production cross sections at LHC with
$\sqrt{s}=14$ TeV for the inoAMSB model with $\tan\beta =10$ and 
$\mu >0$.
}
\label{fig:xs}
\end{center}
\end{figure}

The major visible production reactions consist of 
$pp\to \tg\tg,\ \tg\tq$ and $\tq\tq$ production (here, we take $\tq$ to 
represent generic species of both squarks and anti-squarks).
In the case of inoAMSB models, we expect $m_{\tq}\sim 0.9 m_{\tg}$. 
Strongly interacting sparticle production cross sections 
(at NLO using Prospino\cite{prospino}) are shown versus
$m_{3/2}$ in Fig. \ref{fig:xs} for $\tan\beta =10$, $\mu >0$ and $M_s=M_{GUT}$.
We see that the reactions $pp\to\tq\tq$ and $\tq\tg$ are roughly comparable, 
with $\tq\tg$ production dominating for $m_{3/2}\alt 65$ TeV, and $\tq\tq$ pair production
dominating for higher $m_{3/2}$ values. The $pp\to\tg\tg$ 
production cross section always occurs at much lower rates.
For $m_{\tg}\sim 3$ TeV, corresponding to $m_{3/2}\sim 150$ TeV, the total
hadronic SUSY cross section is around 0.1 fb, which should be around the upper limit of
LHC reach given 100 fb$^{-1}$ of integrated luminosity.

Since sleptons are much lighter than squarks in inoAMSB models, 
we also expect possibly observable rates for slepton pair production.
Pair production rates for $pp\to \te_L^+\te_L^-$, $\te_R^+\te_R^-$ and
$\tnu_{eL}\te_L$ are also shown in Fig. \ref{fig:xs}.
Typically, the LHC reach for direct slepton pair production ranges up to
$m_{\tell}\sim 350$ GeV for 10 fb$^{-1}$\cite{slepton}, corresponding to a 
$m_{3/2}$ value of $\sim 75$ TeV. Thus, LHC reach should be much higher in the
hadronic SUSY production channels.

\subsection{Sparticle decays in inoAMSB models}

Since $m_{\tg}>m_{\tq}$ in inoAMSB models, we will have $\tg\to q\tq$, 
nearly democratically to all squark species. 
The left-squarks will dominantly decay to $wino+q$, and we find
$\tq_L\to q\tw_1$ at $\sim 67\%$, while $\tq_L\to q\tz_1$ at $\sim 33\%$,
all over parameter space. The right-squark decays are simpler.
The $\tq_R$ decays mainly to $bino+q$, so that in the inoAMSB model line, 
we obtain $\tq_R\to\tz_2 q$ at $\sim 97\%$ over almost all parameter space, 
since in this case $\tz_2$ is nearly pure bino-like. 

For the sleptons, the left-sleptons dominantly decay to
$wino+lepton$, so we find $\tell_L\to\ell\tz_1$ at $\sim 33\%$, and 
$\tell_L\to\tw_1\nu_{\ell L}$ at $\sim 67\%$ all over parameter space.
The latter decay mode should be nearly invisible, unless the 
highly ionizing $\tw_1$ track is found in the micro-vertex detector.
The sneutrino decays as $\tnu_{\ell L}\to \tz_1\nu_\ell$ at $\sim 33\%$,
which is again nearly invisible. However, it also decays via
$\tnu_{\ell L}\to \ell\tw_1$ at $\sim 66\%$, which provides a detectable
decay mode for the sneutrinos. The $\te_R$ would like to decay to
$bino+lepton$, but in the case of inoAMSB models, the bino-like
neutralino is too heavy for this decay to occur. In the case of
inoAMSB1 benchmark point, we instead get $\tell_R\to e\tz_1$ at
$\sim 78\%$. Since this decay mode is suppressed, some three body decay 
modes can become comparable. In his case, we find 
$\tell_R^-\to \ell^-\tau^+\ttau_1^-$ at $\sim 13\%$, and 
$\tell_R^-\to\ell^-\tau^-\ttau_1^+$ at $\sim 7\%$.
 
\subsection{LHC collider events for the inoAMSB models}

We use Isajet 7.80\cite{isajet} for the simulation of signal and 
background events at the LHC. A toy detector simulation is employed with
calorimeter cell size
$\Delta\eta\times\Delta\phi=0.05\times 0.05$ and $-5<\eta<5$. The 
hadronic calorimeter (HCAL)
energy resolution is taken to be $80\%/\sqrt{E}+3\%$ for $|\eta|<2.6$ and
forward calorimeter (FCAL) is $100\%/\sqrt{E}+5\%$ for $|\eta|>2.6$. 
The electromagnetic (ECAL) energy resolution
is assumed to be $3\%/\sqrt{E}+0.5\%$. We use the UA1-like jet finding algorithm
GETJET with jet cone size $R=0.4$ and require that $E_T(jet)>50$ GeV and
$|\eta (jet)|<3.0$. Leptons are considered
isolated if they have $p_T(e\ or\ \mu)>20$ GeV and $|\eta|<2.5$ with 
visible activity within a cone of $\Delta R<0.2$ of
$\Sigma E_T^{cells}<5$ GeV. The strict isolation criterion helps reduce
multi-lepton backgrounds from heavy quark ($c\bar c$ and $b\bar{b}$) production.

We identify a hadronic cluster with $E_T>50$ GeV and $|\eta(j)|<1.5$
as a $b$-jet if it contains a $B$ hadron with $p_T(B)>15$ GeV and
$|\eta (B)|<3$ within a cone of $\Delta R<0.5$ about the jet axis. We
adopt a $b$-jet tagging efficiency of 60\%, and assume that
light quark and gluon jets can be mis-tagged as $b$-jets with a
probability $1/150$ for $E_T<100$ GeV, $1/50$ for $E_T>250$ GeV, 
with a linear interpolation for $100$ GeV$<E_T<$ 250 GeV\cite{xt}. 

We have generated 2M events for case inoAMSB1 from Table \ref{tab:BMs}.
In addition, we have generated background events using Isajet for
QCD jet production (jet-types include $g$, $u$, $d$, $s$, $c$ and $b$
quarks) over five $p_T$ ranges as shown in Table \ref{tab:bg}. 
Additional jets are generated via parton showering from the initial and final state
hard scattering subprocesses.
We have also generated backgrounds in the $W+jets$, $Z+jets$, 
$t\bar{t}(172.6)$ and $WW,\ WZ,\ ZZ$ channels at the rates shown in 
the same Table. The $W+jets$ and $Z+jets$ backgrounds
use exact matrix elements for one parton emission, but rely on the 
parton shower for subsequent emissions.

For our initial selection of signal events, we first require the following 
minimal set of cuts labeled ${\bf C1}$:
\bi
\item $n(jets)\ge 2$,
\item $\eslt >max\ (100\ {\rm GeV},0.2M_{eff})$
\item  $E_T(j1,\ j2)>100,\ 50$ GeV,
\item  transverse sphericity $S_T>0.2$,
\ei
where $M_{eff}=\eslt +E_T(j1)+E_T(j2)+E_T(j3)+E_T(j4)$.

\begin{figure}[htbp]
\begin{center}
\includegraphics[angle=0,width=.8\textwidth]{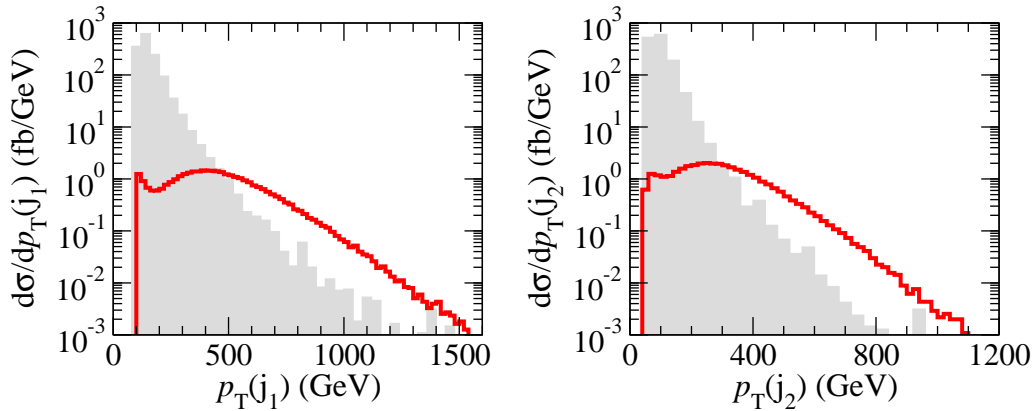}
\caption{Distribution in $p_T$ of {\it a}). the hardest and {\it b}). 
second hardest jets 
from the inoAMSB1 model, and summed SM background (gray histogram), 
for LHC collisions at $\sqrt{s}=14$ TeV.
}
\label{fig:ptj}
\end{center}
\end{figure}

Since sparticle production in inoAMSB models is dominated by $\tq\tg$ and $\tq\tq$ reactions,
followed by $\tq\to q\tz_i$ or $q'\tw_j$, we expect at least two very hard jets in each 
signal event. In Fig. \ref{fig:ptj}, we plot out the distribution in 
{\it a}). hardest and {\it b}). second hardest jet $p_T$ for the signal case inoAMSB1 
along with the summed SM background (denoted by gray histograms).
In the case of $p_T(j_1)$, background is dominant for lower $p_T$ values
$\alt 400$ GeV, while signal emerges from background for higher $p_T$ values.
In the case of $p_T(j_2)$, signal emerges from background already around
250-300 GeV. The rather hard jet $p_T$ distributions are characteristic of
squark pair production, followed by 2-body squark decay into a hard jet.

\begin{figure}[htbp]
\begin{center}
\includegraphics[angle=0,width=0.8\textwidth]{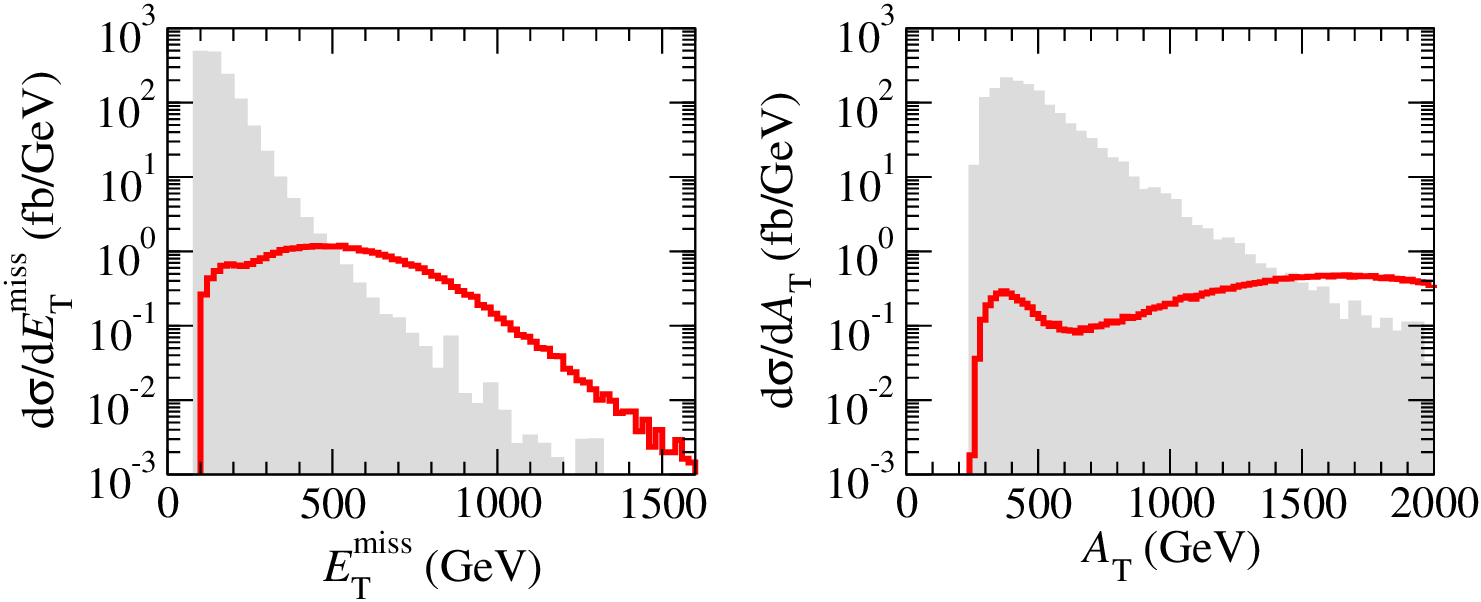}
\caption{Distribution in {\it a}). $\eslt$ and {\it b}). $A_T$ 
from the inoAMSB1 model, and summed SM background (gray histogram), 
for LHC collisions at $\sqrt{s}=14$ TeV.
}
\label{fig:etat}
\end{center}
\end{figure}

In Fig. \ref{fig:etat}, we show the distributions in {\it a}). $\eslt$ 
and {\it b}). $A_T=\sum E_T$ (where the sum extends over all jets and isolated leptons) 
expected from inoAMSB1 along with SM background. In this case, the $\eslt$ distribution 
from SUSY emerges from background at around 400-500 GeV, illustrating the rather
hard $\eslt$ distribution expected from $\tg\tq$ and $\tq\tq$ pair production, followed
by 2-body decays. The $A_T$ signal distribution actually exhibits two components:
a soft peak around 400 GeV which comes from chargino, neutralino and slepton
pair production, and a hard peak at much higher values coming from gluino and squark
pair production. The low peak is buried under background, while the higher peak emerges 
from background at around 1400 GeV.

\begin{figure}[htbp]
\begin{center}
\includegraphics[angle=0,width=0.8\textwidth]{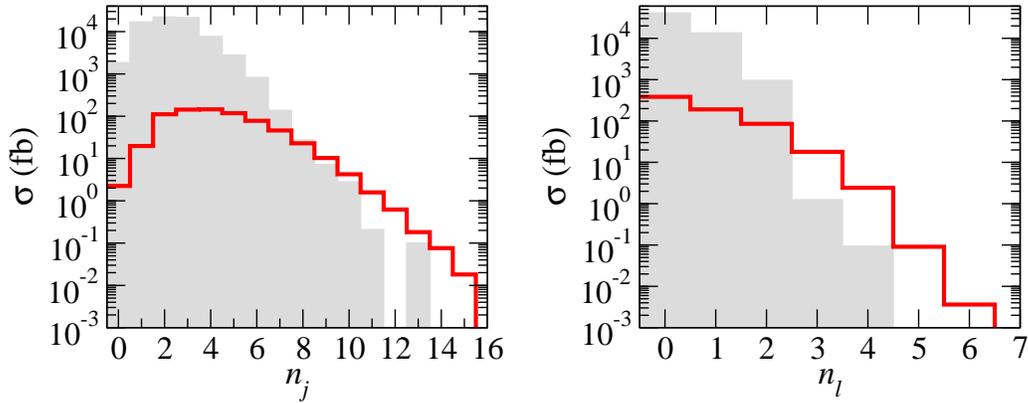}
\caption{Distribution in {\it a}). $n(jets)$ and {\it b}). $n(leptons)$ 
from the inoAMSB1 model, and summed SM background (gray histogram), 
for LHC collisions at $\sqrt{s}=14$ TeV.
}
\label{fig:njnl}
\end{center}
\end{figure}

Fig. \ref{fig:njnl} shows the distribution in {\it a}). jet multiplicity $n_j$ and
{\it b}). isolated lepton (both $e$s and $\mu$s) multiplicity $n_\ell$ 
from the inoAMSB1 benchmark, compared to SM background after C1 cuts. 
While the signal is dominated by $\tq\tq$ and $\tq\tg$ pair production, the 
jet multiplicity actually exhibits a broad peak around $n_j\sim 2-5$. 
Nominally, we would expect dijet dominance from squark pair production. But 
additional jets from cascade decays and initial state radiation help broaden the distribution.
The broadness of the distribution also depends on our jet $E_T$ cut, which requires only that
$E_T(jet)>50$ GeV. 
In the case of isolated lepton multiplicity, we see that
background dominates signal for $n_{\ell}=0,\ 1$ and 2. However, BG drops more
precipitously as $n_\ell$ increases, so that for $n_{\ell}=3$ or 4, 
signal now dominates background\cite{nlep}. 
In these cases, even with minimal cuts, 
an isolated $3\ell+\ge 2$ jets$+\eslt$ signal should 
stand out well above background.

\subsubsection{LHC cascade decay events including HITs:
a smoking gun for models with wino-like neutralinos}

Of course, a distinctive property of models like inoAMSB (and also mAMSB and HCAMSB) 
with a wino-like $\tz_1$ state is that the chargino is very long lived\cite{matchev}: 
of order $\sim 10^{-10}$ sec. Thus, 
once we have obtained cascade decay signal events in any of the multi-jet plus multi-lepton 
plus $\eslt$ channels, we may in addition
look for the presence of a highly-ionizing track (HIT) from the long-lived chargino. 
The presence of HITs in the SUSY collider events would be indictative of models
such as inoAMSB, mAMSB or HCAMSB, where $M_2\ll M_1$ and $M_3$, so that the lightest
neutralino is a nearly pure wino state and where $m_{\tw_1}\simeq m_{\tz_1}$.

\subsection{The reach of LHC in the inoAMSB model line}

We would next like to investigate the reach of the CERN LHC for SUSY in the
inoAMSB context. To this end, we will adopt the inoAMSB model line with variable
$m_{3/2}$ but fixed $\tan\beta =10$ and $\mu >0$. 
The sparticle mass spectra versus $m_{3/2}$ was shown previously in Fig. \ref{fig:m10}

Motivated by the previous signal and background distributions, 
we will require the following cuts $C2$\cite{bbkt}:
\begin{itemize}
\item $n(jets)\ge 2$
\item $S_T>0.2$
\item $E_T(j1),\ E_T(j2),\ \eslt >E_T^c$,
\end{itemize} 
where $E_T^c$ can be variable. Parameter space points with lower sparticle masses
will benefit from lower choices of $E_T^c$, while points with heavier sparticle
masses-- with lower cross sections but higher energy release per event-- will benefit from
higher choices of $E_T^c$.
In addition, in the zero-leptons channel we require 
$30^\circ <\Delta\phi (\vec{E}_T^{miss},{\vec{E}_T}(j_c))<90^\circ$ between the $\vec{E}_T^{miss}$ 
and the nearest jet in transverse opening angle. 
For all isolated leptons $\ell$, we require $p_T(\ell )>20$ GeV.
We separate the signal event channels according to the multiplicity of isolated
leptons: we exhibit the $0\ell$, opposite-sign (OS) dilepton, 
$3\ell$ and $4\ell$ channels. 
Here, we do not here require ``same flavor'' on the OS dilepton events.
We suppress the $1\ell$ and same-sign dilepton $SS$ channels for brevity, and 
because the reach is better in the channels shown.

The resultant cross sections after cuts $C2$ for SM backgrounds along with 
signal point inoAMSB1 are listed in Table \ref{tab:bg} for $E_T^c=100$ GeV.
For each BG channel, we have generated $\sim 2$ million simulated events. 
With the hard cuts $C2$, we are unable to pick up BG cross sections in
some of the multi-lepton channels. 
We will consider a signal to be observable at an assumed value of integrated luminosity 
if {\it i})~the signal to
background ratio, $S/BG \ge 0.1$, {\it ii})~the signal has a minimum of
five events, and {\it iii})~the signal satifies a statistical criterion
$S \ge 5\sqrt{BG}$ (a $5\sigma$ effect).
%
\begin{table}
\begin{center}
\begin{tabular}{lccccc}
\hline
process & $0\ell$ & $OS$ & $SS$ & $3\ell$ & $4\ell$ \\
\hline
QCD($p_T$: 0.05-0.10 TeV) & -- & -- & -- & -- & -- \\
QCD($p_T$: 0.10-0.20 TeV) & 755.1 & -- & -- & -- & -- \\
QCD($p_T$: 0.20-0.40 TeV) & 803.8 & 621.1 & 109.6 & 36.5 & --  \\
QCD($p_T$: 0.40-1.00 TeV) & 209.8 & 304.7 & 72.6 & 29.0 & 2.6 \\
QCD($p_T$: 1.00-2.40 TeV) & 2.2   & 5.3 & 1.7 & 1.5 & 0.2 \\
$t\bar{t}$ & 1721.4 & 732.6 & 273.8 & $ 113.3 $ & $ 6.6$ \\
$W+jets; W\to e,\mu,\tau$ & 527.4 & 22.6  & 8.4 & $ 1.3 $ & $ -- $ \\
$Z+jets; Z\to \tau\bar{\tau},\ \nu s$ & 752.9 & 11.1 & 1.3 & $0.2$ & $ -- $ \\
$WW,ZZ,WZ$ & 3.4 & 0.3 & $0.25$ & $ -- $ & $ -- $ \\
\hline
$summed\ SM\ BG$ & 4776.1 & 1697.8 & 467.7 & $181.9$ & $9.4$ \\
\hline
inoAMSB1 & 112.7 & 85.7 & 27.6 & 36.0 & 7.5 \\
\hline
\end{tabular}
\caption{Estimated SM background cross sections (plus the inoAMSB1 benchmark 
point) in fb for various multi-lepton
plus jets $+\eslt$ topologies after cuts C2 with $E_T^c=100$ GeV.
}
\label{tab:bg}
\end{center}
\end{table}

\begin{figure}[htbp]
\begin{center}
\includegraphics[angle=0,width=0.7\textwidth]{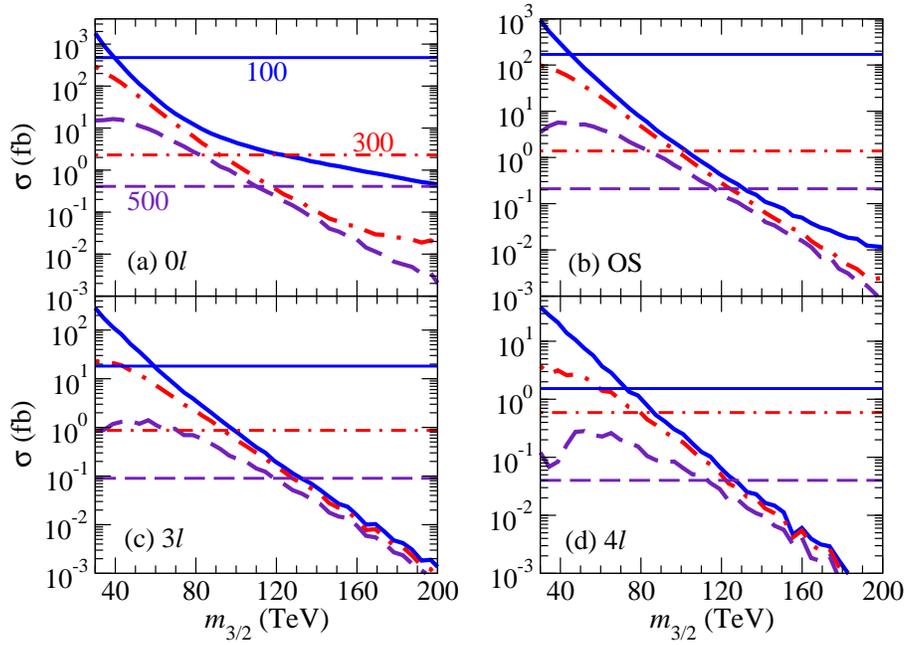}
\caption{Cross section for multi-jet plus $\eslt$
events with {\it a}). $n(\ell) =0$,  {\it b}). OS isolated dileptons
{\it c}). isolated $3\ell$s and {\it d}). isolated $4\ell$s 
at the LHC after cuts $C2$ listed in 
the text with $E_T^c=100$ GeV (blue solid), $E_T^c=300$ GeV (red dot-dashed) and
$E_T^c=500$ GeV (purple dashes), versus $m_{3/2}$, 
from the inoAMSB model line points 
with $\tan\beta =10$ and $\mu >0$.
We also list the 100 fb$^{-1}$ $5\sigma$, 5 event, $S>0.1\ BG$ 
limit with the horizontal lines.  
}
\label{fig:reach}
\end{center}
\end{figure}

Using the above criteria, the 100 fb$^{-1}$ reach of the LHC can be computed
for each signal channel. 
In Fig. \ref{fig:reach}, we show the signal rates versus $m_{3/2}$ 
for the inoAMSB model line for $E_T^c=100$ (solid blue), 300 (dot-dash red) and 500 GeV
(dashed purple).
The 100 fb$^{-1}$ LHC reach is denoted by the 
horizontal lines for each $E_T^c$ value.
From frame {\it a})., for the multi-jet$+\eslt+0\ell$ signal, 
we see the LHC reach in the $0\ell$ channel extends to
$m_{3/2}\sim 40,\ 93$ and 110 TeV for $E_T^c=100,\ 300$ and 500 GeV,
respectively, for the inoAMSB model line. 
This corresponds to a reach in
$m_{\tg}$ of 1.1, 2.0 and 2.4 TeV. 

Frames {\it b})., {\it c}). and {\it d}). show the reach in the
multi-jet$+\eslt+$ $OS$, $3\ell$ and $4\ell$ channels, respectively.
While the reach is qualitatively similar in all channels, the best reach
comes from the $3\ell$ channel, where the 100 fb$^{-1}$ LHC can detect
inoAMSB models up to $m_{3/2}\sim 118$ TeV (corresponding to a reach in
$m_{\tg}$ of 2.6 TeV), using $E_T^c=500$ GeV. The 100 fb$^{-1}$ LHC 
reach for all cases 
is summarized in Table \ref{tab:reach}.
%
\begin{table}
\begin{center}
\begin{tabular}{lcccc}
\hline
$E_T^c$ (GeV) & $0\ell$ & $OS$ & $3\ell$ & $4\ell$  \\
\hline
100 & $40$ & $57$ &  $60$ & $75$ \\
300 & $93$ & $95$ &  $98$ & $80$ \\
500 & $110$ & $115$ & $118$ & $110$ \\
\hline
\end{tabular}
\caption{Estimated reach of 100 fb$^{-1}$ LHC for $m_{3/2}$ (TeV) in 
the inoAMSB  model line in various signal channels.
}
\label{tab:reach}
\end{center}
\end{table}

\subsubsection{Cascade decays including HITs plus a multi-bump 
$m(\ell^+\ell^-)$ distribution: a smoking gun for inoAMSB models}

Next, we examine the distribution in $m(\ell^+\ell^- )$ for cascade decay events
containing: $\ge 2$ high $p_T$ jets, large $\eslt$ and a pair of same flavor/opposite-sign 
(SF/OS) dileptons.
This distribution has for long been touted
as being very useful as a starting point for reconstructing sparticle masses
in SUSY cascade decay events, because it may contain a kinematic mass edge from
$\tz_2\to\tell^\pm\ell^\mp$ or $\tz_2\to\ell^+\ell^-\tz_1$ decays. 
In the case of the inoAMSB1 benchmark model, where $m_{\tell_{L,R}}<m_{\tz_2}$-- and a substantial
mass gap between $m_{\tell_L}$ and $m_{\tell_R}$ is featured-- 
we expect {\it two} distinct, well-separated
mass edges: one from $\tz_2\to\tell_L\ell$ and one from $\tz_2\to\tell_R\ell$ decays.
In addition, a peak at $m(\ell^+\ell^-)\sim M_Z$ is expected, since real $Z$ bosons can 
be emitted from cascade decays including $\tz_3\to Z\tz_1$, $\tz_4\to Z\tz_1$ and
$\tw_2\to Z\tw_1$ (in the case of benchmark model inoAMSB1, these decays occur
with branching fractions 25\%, 6\% and 29\%, respectively). 

\begin{figure}[htbp]
\begin{center}
\vspace{.5in}
\includegraphics[angle=-90,width=0.9\textwidth]{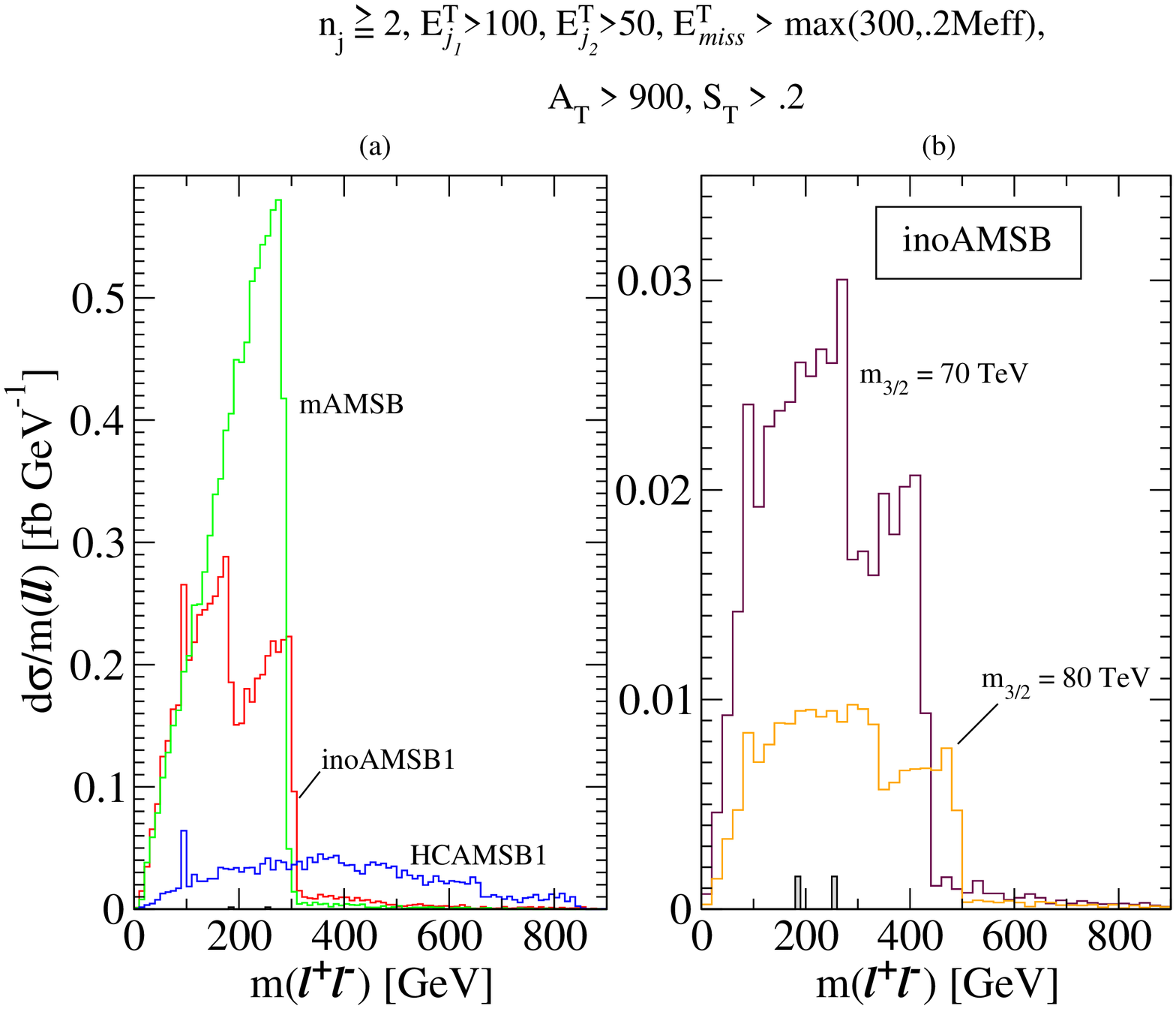}
\caption{Invariant mass distribution for SF/OS dileptons
from {\it a}). mAMSB1, HCAMSB1 and inoAMSB1 after requiring cut set 
$C1$ plus $\eslt>300$ GeV and $A_T>900$ GeV. In frame {\it b}), we show the same
distribution, except taking inoAMSB with $m_{3/2}=70$ and 80 TeV.
}
\label{fig:mll}
\end{center}
\end{figure}

In Fig. \ref{fig:mll}, we show the $m(\ell^+\ell^-)$ distribution from inoAMSB1 
(red histogram) in frame
{\it a}). Here, we require cuts $C1$, along with $\eslt >300$ GeV and $A_T>900$ GeV,
which completely suppresses SM backgrounds.
Indeed, we see clearly a $Z$ boson peak at $M_Z$, along with two distinct mass edges
occuring at $m(\ell^+\ell^- )=
m_{\tz_2}\sqrt{1-\frac{m_{\tell}^2}{m_{\tz_2}^2}}\sqrt{1-\frac{m_{\tz_1}^2}{m_{\tell}^2}}= 182$ GeV, and 304 GeV.
The 182 GeV edge comes from $\tz_2$ decays through $\tell_R$, while the 304 GeV edge comes
from $\tz_2$ decays through $\tell_L$. 
We also show the same distribution for the mAMSB1 (green) and HCAMSB1 (blue) cases from Table \ref{tab:BMs}.
The mAMSB plot contains two mass edges as well. However, since in mAMSB we expect
$m_{\tell_L}\simeq m_{\tell_R}$, these edges nearly overlap, and are essentially indistinguishable.
In the case of HCAMSB models, the bino-like neutralino is the $\tz_4$ and is quite heavy, 
while $\tz_2$ and $\tz_3$ are mainly higgsino-like. The higgsino-like states decay strongly to vector bosons,
as does $\tw_2$, giving rise to a continuum $m(\ell^+\ell^-)$ distribution which contains a 
$Z$ peak\cite{hcamsb}. 
Thus, while the presence of SUSY cascade decay events at LHC containing HITs would point to
AMSB-like models, the different $m(\ell^+\ell^-)$ distributions which are expected
would allow one to differentiate between the mAMSB, HCAMSB and inoAMSB cases!

In frame {\it b})., we show inoAMSB models with $m_{3/2}=70$ and 80 GeV. These distributions
also show the expected double edge plus $Z$ peak structure that was found for inoAMSB1, 
although now the mass edges have migrated to higher $m(\ell^+\ell^-)$ values. 

\section{Discussion and conclusions}
\label{sec:conclude}

In this paper, we have examined the phenomenology of supersymmetric models
with the boundary conditions $m_0\sim A_0\sim 0$ at $M_{GUT}$, while
gaugino masses assume the form as given in AMSB. We call this model
gaugino-AMSB, or inoAMSB or short. Such boundary conditions can arise in
type IIB string models with flux compactifications. They are very compelling in
that off-diagonal flavor violating and also $CP$ violating terms are highly suppressed,
as in the case of no-scale supergravity or gaugino-mediated SUSY breaking models. 
However, since gaugino masses assume the AMSB form at $M_{GUT}$, the large
$U(1)_Y$ gaugino mass $M_1$ pulls slepton masses to large enough values through
renormalization group evolution that one avoids charged LSPs (as in NS or 
inoMSB model) or tachyonic sleptons (as in pure AMSB models).

The expected sparticle mass spectrum is very distinctive. Like mAMSB and HCAMSB, 
we expect a wino-like lightest neutralino $\tz_1$, and a quasi-stable chargino
$\tw_1$ which could leave observable highly ionizing tracks in a collider
detector. The spectrum is unlike mAMSB in that a large mass splitting is expected
between left- and right- sleptons.
We also investigated what happens if the string scale $M_s$ is much lower than $M_{GUT}$.
In this case, the entire spectrum become somewhat expanded, and  if $M_s\alt 10^{14}$ GeV, 
then the left-sneutrino becomes the LSP, which is excluded by double beta decay experiments.

We also investigated in detail some aspects of LHC collider signatures.
Since $m_{\tq}<m_{\tg}$ in inoAMSB models, we expect dominant $\tq\tq$ and $\tq\tg$ 
production at LHC, followed by 2-body $\tq$ and $\tg$ decays. This leads to
collider events containing at least two very high $p_T$ jets plus $\eslt$ as is
indicative from squark pair production.

While squark and gluino cascade decay events should be easily seen at LHC (provided
$m_{3/2}\alt 110$ TeV), the signal events should all contain visible HITs, which
would point to a model with $m_{\tw_1}\simeq m_{\tz_1}$, as occurs in anomaly-mediation
where $M_2<M_1,\ M_3$ at the weak scale.
We find an LHC reach for 100 fb$^{-1}$ of integrated luminosity out to 
$m_{3/2}\sim 118$ TeV, corresponding to a reach in $m_{\tg}$ of about 2.6 TeV.

We also find that the invariant mass distribution of SF/OS dilepton pairs should have
a distinctive two-bump structure that is indicative of neutralino decays through
both left- and right- sleptons with a large slepton mass splitting. This distribution would help
distinguish inoAMSB models from HCAMSB, where a continuum plus a $Z$-bump distribution is expected,
or from mAMSB, where the two mass edges (present only if $m_0$ is small enough that
$m_{\tell_L}$ and $m_{\tell_R}$ are lighter than $m_{\tz_2}$) would be very close together,
and probably not resolvable.

\section*{Acknowledgments}
This work was supported in part by the U.S.~Department of Energy. 
SdA and KG are supported in part by the United States Department of Energy under grant DE-FG02-91-ER-40672. 
%

%

\begin{thebibliography}{99}
\small
%
\bibitem{Grana:2005jc}
  M.~Grana,
  Phys.\ Rept.\  {\bf 423}, 91 (2006)
  [arXiv:hep-th/0509003].
  %
\bibitem{Douglas:2006es}
  M.~R.~Douglas and S.~Kachru,
  Rev.\ Mod.\ Phys.\  {\bf 79}, 733 (2007)
  [arXiv:hep-th/0610102].
  
%
\bibitem{Kachru:2003aw} S. Kachru, R. Kallosh, A. Linde and S. Trivedi,
\prd{68}{2003}{046005}.
%
\bibitem{Balasubramanian:2005zx} V. Balasubramanian, P. Berglund, J. P. Conlon
 and F. Quevedo, \jhep{03}{2005}{007}.
%
\bibitem{amsb} L. Randall and R. Sundrum, \npb{557}{1999}{79}; 
G. Giudice, M. Luty, H. Murayama and R. Rattazzi, \jhep{12}{1998}{027}.
%
\bibitem{deAlwis:2008aq} S. P. de Alwis, \prd{77}{2008}{105020} arXiv:0801.0578.
%
\bibitem{wss} See
H.~Baer and X.~Tata, {\it Weak Scale Supersymmetry: From                        
Superfields to Scattering Events},
(Cambridge University Press, 2006)
%
\bibitem{noscale} 
J.~R.~Ellis, C.~Kounnas and D.~V.~Nanopoulos,
  Nucl.\ Phys.\  B {\bf 247} (1984) 373;
J.~R.~Ellis, A.~B.~Lahanas, D.~V.~Nanopoulos and K.~Tamvakis,
  Phys.\ Lett.\  B {\bf 134} (1984) 429;
G.~A.~Diamandis, J.~R.~Ellis, A.~B.~Lahanas and D.~V.~Nanopoulos,
  Phys.\ Lett.\  B {\bf 173} (1986) 303;
For a review, see A. Lahanas and D. V. Nanopoulos, 
\prep{145}{1987}{1}.
%
\bibitem{inomsb} D. E. Kaplan, G. D. Kribs and M. Schmaltz, 
\prd{62}{2000}{035010}; Z. Chacko, M. A. Luty, A. E. Nelson and E. Ponton, 
\jhep{01}{2000}{003}.
%
\bibitem{masiero} F. Gabbiani, E. Gabrielli, A. Masiero and L. Silvestrini, 
\npb{477}{1996}{321}.
%
\bibitem{ss} M. Schmaltz and W. Skiba, \prd{62}{2000}{095005} and 
\prd{62}{2000}{095004}; H. Baer, A. Belyaev, T. Krupovnickas
and X. Tata, \prd{65}{2002}{075024};
A. De Simone, J. J. Fan, M. Schmaltz and W. Skiba,
\prd{78}{2008}{095010}.
%
\bibitem{derm_mafi} R. Dermisek and A. Mafi, \prd{65}{2002}{055002}; for associated 
phenomenology, see H.~Baer, C.~Balazs, A.~Belyaev, R.~Dermisek, A.~Mafi and A.~Mustafayev,
\jhep{0205}{2002}{061} and Nucl.\ Instrum.\ Meth.\  A {\bf 502} (2003) 560.
%
\bibitem{hcamsb} R. Dermisek, H. Verlinde and L-T. Wang, \prl{100}{2008}{131804}; 
H. Baer. R. Dermisek, S. Rajagopalan and H. Summy, \jhep{0910}{2009}{078}.
%
\bibitem{Wess:1992cp} J. Wess and J. Bagger, {\it Supersymetry and Supergravity}, (Princeton University Press, 1992).
%
\bibitem{Gates:1983nr} S. J. Gates, M. T. Grisaru, M. Rocek  and W. Siegel,
{\it Superspace, or one thousand and one lessons in supersymmetry}, 
Front. Phys. {\bf 58} (1983) 1.
%
\bibitem{deAlwis:2008kt} S. P. de Alwis, \jhep{0903}{2009}{023} arXiv:0806.2672.
%
\bibitem{Blumenhagen:2007sm}
  R.~Blumenhagen, S.~Moster and E.~Plauschinn,
  JHEP {\bf 0801}, 058 (2008)
  [arXiv:0711.3389 [hep-th]].
   %
\bibitem{Blumenhagen:2009gk} R. Blumenhagen, S.  Moster, and E. Plauschinn,
\jhep{0801}{2008}{058}.
%
\bibitem{deAlwis:2009tp} S. P. de Alwis ``Classical and Quantum SUSY Breaking Effects in IIB Local Models,''  arXiv:0912.2950.
%
\bibitem{Bagger:1999rd}J. Bagger, T. Moroi, and
 E. Poppitz, \jhep{04}{2000}{009},   hep-th/9911029.
 %
\bibitem{Kaplunovsky:1994fg} V. Kaplunovsky and J. Louis,
\npb{422}{1994}{57}.
%
\bibitem{ArkaniHamed:1997mj} N. Arkani-Hamed and H. Murayama,
\jhep{0006}{2000}{030}.
%
%
\bibitem{moroi} K. Kohri, T. Moroi and A. Yotsuyanagi, \prd{73}{2006}{123511};
for an update, see
M. Kawasaki, K. Kohri, T. Moroi and A. Yotsuyanagi, \prd{78}{2008}{065011};
see also J.~Pradler and F.~D.~Steffen, \plb{648}{2007}{224}.
%
\bibitem{isajet} F. Paige, S. Protopopescu, H. Baer and X. Tata, \hepph{0312045}.
%
\bibitem{hh} H. E. Haber, R. Hempfling and A. Hoang, \zpc{75}{1997}{539}.
%
\bibitem{pbmz} D. Pierce, J. Bagger, K. Matchev and R. Zhang, \npb{491}{1997}{3}.
%
\bibitem{kraml} G. Belanger, S. Kraml and A. Pukhov, \prd{72}{2005}{015003}.
%
\bibitem{brown} H.N. Brown {\it et al.} (Muon $g-2$ collaboration), 
\prl{86}{2001}{2227}.
%
\bibitem{isagm2} H. Baer, C. Balazs, J. Ferrandis and X. Tata, \prd{64}{2001}{035004}.
%
\bibitem{amsb_lhc} T. Gherghetta, G. Giudice and J. Wells, \npb{559}{1999}{27};
J. L. Feng, T. Moroi, L. Randall, M. Strassler and S. Su, \prl{83}{1999}{1731};
J. L. Feng and T. Moroi, \prd{61}{2000}{095004}; F. Paige and J. Wells, \hepph{0001249};
A. Datta and K. Huitu, \prd{67}{2003}{115006};
S. Asai, T. Moroi, K. Nishihara and T. T. Yanagida, \plb{653}{2007}{81};
S. Asai, T. Moroi and T. T. Yanagida, \plb{664}{2008}{185};
H. Baer, J. K. Mizukoshi and X. Tata, \plb{488}{2000}{367};
A. J. Barr, C. Lester, M. Parker, B. Allanach and P. Richardson,
\jhep{0303}{2003}{045}.
%
\bibitem{lepw1lim} LEPSUSYWG, note LEPSUSYWG/02-04.1.
%
\bibitem{sneu_dm} 
S. P. Ahlen {\it et al.}, \plb{195}{1987}{603};
D. Caldwell {\it et al.}, \prl{61}{1988}{510} and \prl{65}{1990}{1305};
D. Reusser {\it et al.}, \plb{235}{1991}{143};
T. Falk, K. Olive and M. Srednicki, \plb{339}{1994}{248}.
%
\bibitem{isabsg} H. Baer and M. Brhlik, \prd{55}{1997}{3201}.
%
\bibitem{misiak} M.~Misiak {\it et al.}, \prl{98}{2007}{022002}.
%
\bibitem{bsg_ex} E.~Barberio {\it et al.} (Heavy Flavor Averaging Group), 
\hepex{0603003}.
%
\bibitem{bbct} H. Baer, M. Brhlik, D. Castano and X. Tata, \prd{58}{1998}{015007}.
%
\bibitem{randallmoroi} T. Moroi and L. Randall, \npb{570}{2000}{455}.
%
\bibitem{axion} L. F. Abbott and P. Sikivie, \plb{120}{1983}{133};
J. Preskill, M. Wise and F. Wilczek, \plb{120}{1983}{127};
M. Dine and W. Fischler, \plb{120}{1983}{137};
M. Turner, \prd{33}{1986}{889}.
%
\bibitem{axino} K. Rajagopal, M. Turner  and F. Wilczek, \npb{358}{1991}{447};
 L. Covi, J. E. Kim and L. Roszkowski, \prl{82}{1999}{4180}; 
L. Covi, H. B. Kim, J. E. Kim and L. Roszkowski, \jhep{0105}{2001}{033};
for recent  reviews, see L. Covi and J. E. Kim, arXiv:0902.0769 and 
F. Steffen, \epjc{59}{2009}{557}.
%
\bibitem{mix} H. Baer, A. Box and H. Summy, \jhep{0908}{2009}{080};
H. Baer and A. Box, arXiv:0910.0333 (2009).
%
\bibitem{prospino}
Prospino, by W.~Beenakker, R.~Hopker and M.~Spira,
  arXiv:hep-ph/9611232.
%
\bibitem{slepton} H. Baer, C. H. Chen, F. Paige and X. Tata, \prd{49}{1994}{3283}.
%
\bibitem{xt} R. Kadala, J. K. Mizukoshi and X. Tata, \epjc{56}{2008}{511}.
%
\bibitem{nlep} H. Baer, H. Prosper and H. Summy, \prd{77}{2008}{055017};
H. Baer, A. Lessa and H. Summy, \plb{674}{2009}{49};
H. Baer, V. Barger, A. Lessa and X. Tata, \jhep{0909}{2009}{063}.
%
\bibitem{matchev} H. C. Cheng, B. Dobrescu and K. Matchev, \npb{543}{1999}{47}.
%
\bibitem{bbkt} H. Baer, C. H. Chen, F. Paige and X. Tata, \prd{52}{1995}{2746}
and \prd{53}{1996}{6241}; 
H. Baer, A. Belyaev, T. Krupovnickas and X. Tata, 
\prd{65}{2002}{075024}.
%
%
\end{thebibliography}
\end{document}